\documentclass{article}
\usepackage{arxiv}

\usepackage{amssymb}
\usepackage{amsmath}

\usepackage{pgfplots}
\pgfplotsset{compat=1.18}
\usepgfplotslibrary{external}
\usepackage{tikz}
\usetikzlibrary{positioning, shapes.geometric, arrows, arrows.meta, shapes.multipart, shapes.misc, shapes.symbols, backgrounds, fit, calc, decorations.pathmorphing, mindmap, patterns, graphs}

\tikzstyle{node} = [minimum size=0.7cm, rectangle]
\tikzstyle{box} = [node, draw=black]
\tikzstyle{rounded box} = [box, rounded corners]
\tikzstyle{if} = [diamond, draw=black, aspect=2]
\tikzstyle{helper} = [
    node,
]
\tikzstyle{stack} =  [
    fill = white,
    append after command={
      \pgfextra{\foreach \y in {2,1} {
        \node[box, right=0cm of \tikzlastnode.west, xshift=0.1cm*\y, yshift=0.1cm*\y, text=white, fill=white] {#1};
      }}
    }
  ] \usepackage{makecell}
\usepackage{tabularx}
\usepackage{enumitem}
\usepackage[colorinlistoftodos]{todonotes}
\newlist{questions}{enumerate}{2}
\setlist[questions,1]{label=\textbf{RQ\arabic*},ref=RQ\arabic*}
\setlist[questions,2]{label=\textbf{RQ\arabic*.\arabic*},ref=\thequestionsi.\arabic*}
\usepackage{dirtytalk}
\usepackage{hyperref}
\usepackage{booktabs}
\usepackage{float}
\usepackage{natbib}

\usepackage{subcaption}

\pgfkeys{
  /paper/.cd,
  19/.initial = {19_ahmadDynamicVMProvisioning2021},
  35/.initial = {35_mohanNoSQLDataModel2016},
  49/.initial = {49_kashlevSystemArchitectureRunning2014},
  83/.initial = {83_nascimentoIncrementalReinforcementLearning2021},
  119/.initial = {119_bousselmi_bi_objective},
  126/.initial = {126_simicBigDataBPMN2023},
  133/.initial = {133_chenBigDataProcessing2020},
  182/.initial = {182_aseman-manzarCostAwareResourceRecommendation2023},
  194/.initial = {194_dettiCWLPLASTaskWorkflows2023},
  224/.initial = {224_makhlouf2019data},
  229/.initial = {229_zhangDataintensiveWorkflowScheduling2023},
  291/.initial = {291_ikken_2018},
  322/.initial = {322_kamburugamuveHPTMTOperatorBasedArchitecture2021},
  326/.initial = {326_mousavimojabICATSSchedulingBig2019},
  334/.initial = {334_quinnImplicationsAlternativeServerless2021},
  373/.initial = {373_changxin_lpod},
  379/.initial = {379_mehranMatchingbasedSchedulingAsynchronous2022},
  424/.initial = {424_fanOptimizingDataRegeneration2024},
  426/.initial = {426_wuOptimizingPerformanceBig2016},
  431/.initial = {431_tudoranOverFlowMultiSiteAware2016},
  442/.initial = {442_shuPerformanceOptimizationHadoop2017},
  449/.initial = {449_chenPISCESOptimizingMultiJob2019},
  450/.initial = {450_esteves_2014a},
  474/.initial = {474_muller2014pydron},
  487/.initial = {487_tranReducingTailLatencies2018},
  511/.initial = {511_kllapiScheduleOptimizationData2011},
  514/.initial = {514_aravind_schedulingBigData},
  515/.initial = {515_ebrahimiSchedulingBigData2018},
  554/.initial = {554_yeStorageAwareTaskScheduling2018},
  579/.initial = {579_kacsukFlowbsterCloudOrientedWorkflow2018},
  593/.initial = {593_bugingoDecompositionBasedMultiobjective2021},
  604/.initial = {604_chenTransformationBasedStreamingWorkflow2019},
  623/.initial = {623_hamid_workflowScheduling},
}

\pgfkeys{
  /paperids/.cd,
  19_ahmadDynamicVMProvisioning2021/.initial = {A1},
  182_aseman-manzarCostAwareResourceRecommendation2023/.initial = {A2},
  373_changxin_lpod/.initial = {A3},
  119_bousselmi_bi_objective/.initial = {A4},
  593_bugingoDecompositionBasedMultiobjective2021/.initial = {A5},
  133_chenBigDataProcessing2020/.initial = {A6},
  449_chenPISCESOptimizingMultiJob2019/.initial = {A7},
  604_chenTransformationBasedStreamingWorkflow2019/.initial = {A8},
  194_dettiCWLPLASTaskWorkflows2023/.initial = {A9},
  515_ebrahimiSchedulingBigData2018/.initial = {A10},
  450_esteves_2014a/.initial = {A11},
  424_fanOptimizingDataRegeneration2024/.initial = {A12},
  291_ikken_2018/.initial = {A13},
  579_kacsukFlowbsterCloudOrientedWorkflow2018/.initial = {A14},
  322_kamburugamuveHPTMTOperatorBasedArchitecture2021/.initial = {A15},
  49_kashlevSystemArchitectureRunning2014/.initial = {A16},
  511_kllapiScheduleOptimizationData2011/.initial = {A17},
  224_makhlouf2019data/.initial = {A18},
  379_mehranMatchingbasedSchedulingAsynchronous2022/.initial = {A19},
  35_mohanNoSQLDataModel2016/.initial = {A20},
  514_aravind_schedulingBigData/.initial = {A21},
  326_mousavimojabICATSSchedulingBig2019/.initial = {A22},
  474_muller2014pydron/.initial = {A23},
  83_nascimentoIncrementalReinforcementLearning2021/.initial = {A24},
  334_quinnImplicationsAlternativeServerless2021/.initial = {A25},
  623_hamid_workflowScheduling/.initial = {A26},
  442_shuPerformanceOptimizationHadoop2017/.initial = {A27},
  126_simicBigDataBPMN2023/.initial = {A28},
  487_tranReducingTailLatencies2018/.initial = {A29},
  431_tudoranOverFlowMultiSiteAware2016/.initial = {A30},
  426_wuOptimizingPerformanceBig2016/.initial = {A31},
  554_yeStorageAwareTaskScheduling2018/.initial = {A32},
  229_zhangDataintensiveWorkflowScheduling2023/.initial = {A33},
}

\newcommand{\citepaperalias}[1]{%
  \hyperref[\pgfkeysvalueof{/paper/#1}]{\pgfkeysvalueof{/paperids/\pgfkeysvalueof{/paper/#1}}}%
} %

\newcommand{\citea}[1]{\citet{#1}}

\providecommand{\Description}[1]{}

\title{Optimization Opportunities for Cloud-Based Data Pipeline Infrastructures} %

\newif\ifuniqueAffiliation
\uniqueAffiliationtrue

\ifuniqueAffiliation %
\author{\href{https://orcid.org/0009-0001-0205-8850}{\includegraphics[scale=0.06]{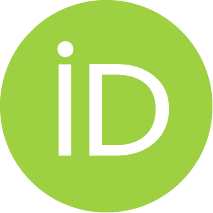}\hspace{1mm}Johannes~Jablonski} \\
	Friedrich-Alexander-Universität Erlangen-Nürnberg\\
	Erlangen, Germany \\
	\texttt{johannes.jablonski@fau.de} \\
	\And
	\href{https://orcid.org/0000-0001-9060-7938}{\includegraphics[scale=0.06]{orcid.pdf}\hspace{1mm}Georg-Daniel~Schwarz} \\
	Friedrich-Alexander-Universität Erlangen-Nürnberg\\
	Erlangen, Germany \\
	\texttt{georg.schwarz@fau.de} \\
    \And
    \href{https://orcid.org/0000-0002-4236-2689}{\includegraphics[scale=0.06]{orcid.pdf}\hspace{1mm}Philip~Heltweg} \\
    	Friedrich-Alexander-Universität Erlangen-Nürnberg\\
    	Erlangen, Germany \\
    	\texttt{philip@heltweg.org} \\
        \And
        \href{https://orcid.org/0000-0002-8139-5600}{\includegraphics[scale=0.06]{orcid.pdf}\hspace{1mm}Dirk~Riehle} \\
	Friedrich-Alexander-Universität Erlangen-Nürnberg\\
	Erlangen, Germany \\
	\texttt{dirk@riehle.org} \\
}
\else
\usepackage{authblk}

\newbox{\orcid}\sbox{\orcid}{\includegraphics[scale=0.06]{orcid.pdf}} 
\author[1]{%
	\href{https://orcid.org/0000-0000-0000-0000}{\usebox{\orcid}\hspace{1mm}David S.~Hippocampus\thanks{\texttt{hippo@cs.cranberry-lemon.edu}}}%
}
\author[1,2]{%
	\href{https://orcid.org/0000-0000-0000-0000}{\usebox{\orcid}\hspace{1mm}Elias D.~Striatum\thanks{\texttt{stariate@ee.mount-sheikh.edu}}}%
}
\affil[1]{Department of Computer Science, Cranberry-Lemon University, Pittsburgh, PA 15213}
\affil[2]{Department of Electrical Engineering, Mount-Sheikh University, Santa Narimana, Levand}
\fi

\renewcommand{\headeright}{Preprint}
\renewcommand{\undertitle}{Preprint}

\hypersetup{
pdftitle={Optimization Opportunities for Cloud-Based Data Pipeline Infrastructures},
pdfauthor={Johannes~Jablonski, Georg-Daniel~Schwarz, Philip~Heltweg, Dirk~Riehle},
}

\begin{document}
\maketitle

\begin{abstract}
Cloud infrastructure supports the efficient operation of data pipelines regarding requirements like cost, speed, and resource utilization.
We present an integrated view of optimization opportunities for cloud-based data pipelines by conducting a systematic review of existing literature on optimization approaches to cloud infrastructure performance for data pipelines.
Our study contributes a theory of optimization goals like minimizing cost, reducing execution time, and cost-makespan trade-offs, consisting of dimensions such as single vs. multi-cloud, batch vs. stream processing, etc.
We highlight gaps in primary research, including the underexploration of multi-tenant environments and lack of industry evaluation, and suggest directions for future research.
\end{abstract}


\section{Introduction}

Data is essential, in research and practice, for academia and industry, to gain knowledge.
To derive knowledge from raw data, it usually needs to be prepared for the analysis that delivers the desired knowledge.
The process of improving data for use in specific projects is called data engineering. While it can be done manually, the implementation of automated data pipelines is a common approach when large amounts of data have to be handled or repeated execution is important. These data pipelines consist of a series of connected processing steps, typically modeled as a directed acyclic graph (DAG). Different styles of data pipelines exist with one of the most common being ETL-pipelines, so called because they first extract data, then transform it to remove errors or improve data quality before finally loading the data into a format suitable for the analysis task at hand. 

Processing large amounts of data with data pipelines leads to a high demand of computing resources, making cloud environments an interesting execution environment. Going to the cloud \say{[...] can provide elastic services and
data-intensive analysis for end-users [...]. Therefore, users can be empowered with seemingly unlimited resources without building new computing infrastructures.} \citep{ren_cloud_2019}.
Several case studies such as \citea{inibhunu2021decupled} and \citea{BYRNE2021100447} indicated the cloud being the natural next step of deploying and operating pipelines.

While the cloud promises significant advantages, the specific objectives that users prioritize for their data pipelines remain unclear. 
For instance, minimizing costs and maximizing execution speed often involve trade-offs, as these goals are typically in tension with one another \citep{511_kllapiScheduleOptimizationData2011}.
In this study, we examined the objectives pursued by data engineers deploying data pipelines in the cloud and analyzed the strategies they employ to achieve these objectives.
\textbf{Our perspective is that of a service provider for cloud-based data pipelines} aiming to better understand user needs and to learn from their optimization strategies, thereby \textbf{enabling the development of more effective and user-aligned data pipeline infrastructure}.

Since cloud-based data pipeline services are typically designed as domain-agnostic and general-purpose offerings, we restrict our focus to implementations that are not tied to specific domains or tools. %
Specifically, we focus on automated optimization opportunities during the execution phase of data pipelines.
While users may provide hints to the pipeline service during the design phase, this phase is primarily driven by manual optimizations and remains outside the scope of service providers.
Our study concentrates on explicitly programmatic approaches that service providers can deploy to enhance execution performance.

Our analysis centers on public clouds but also considers private or hybrid clouds that exhibit characteristics similar to public clouds, such as constrained availability of combinations of different computing resources. 
By contrast, edge, and fog computing paradigms -- which relocate resources closer to users to reduce communication latencies \citep{ren_cloud_2019} -- are out of scope for this study, as they do not align with the typical operational scope of cloud service providers.

To address these goals, we formulate our research questions as follows:

\begin{questions}
    \item How can a pipeline service provider optimize the cloud-based execution of multiple application and domain-independent data pipelines? \label{rq:main}
    \begin{questions}
        \item What optimization goals do existing data pipeline services aim for? \label{rq:goals}
        \item What are well-working optimization techniques for specific data pipeline use cases in the cloud context? \label{rq:techniques}
    \end{questions}
\end{questions}

To achieve these objectives, we conducted a systematic review based on \citet{kitchenham_guidelines_2007}.
This review provides an overview of the current state of the field and can also serve as a reference for people interested in specific optimization goals or methods.
The following key contributions are made by this study:

\begin{itemize}
    \item It provides a systematic analysis of the current state of research on cloud-based data pipeline systems.
    \item It provides a reference for already successfully applied optimizations.
    \item It provides information about research gaps for future studies in the area of data pipelines.
    \item It introduces a theory by synthesizing optimization opportunities into a concept matrix.
    \item It derives connections between data pipeline optimization goals and solutions.
\end{itemize}

Our systematic literature review, detailed in \autoref{sec:research_design}, is supported by a concept matrix.
\autoref{sec:results} presents key findings on optimization goals, contexts, and methods, while \autoref{sec:discussion} interprets these results and identifies areas for future research.
Finally, \autoref{sec:limitations} discusses study limitations, and \autoref{sec:conclusion} summarizes our contributions and suggests future directions.

\section{Related Work} \label{sec:related_work}
The field of data pipelines was extensively studied, with foundational research establishing a robust understanding of their principles and applications. 
For instance, foundational work such as \citea{ritchie1984unix} introduced the mental model of data pipelines: pipes and filters. 
They connected command-line programs (filters) using a simple inter-process communication mechanism (pipes).
The same design approach can still be found in many data pipeline implementations.

\citea{abadi2003aurora} embedded this mental model into Aurora, their architecture for data stream processing. 
They designed the system by connecting operator boxes (e.g., filters, selections, aggregates, splits, unions) into a directed acyclic graph (DAG), allowing for the composition of modular stream processing tasks. 
Aurora efficiently distributes workloads across compute nodes while focusing on Quality of Service (QoS) by monitoring metrics such as response times and tuple drops.
The system incorporates load-shedding mechanisms to handle overload scenarios, ensuring that critical workloads maintain acceptable QoS levels. 
Additionally, it optimizes processing through dynamic scheduling, adapting to changes in workload and resource availability.

Numerous studies explored optimization techniques to enhance the efficiency of data pipelines. For example, \citea{zaharia2010delay} introduced delay scheduling, a technique that prioritizes task locality to reduce data transfer times and improve performance in distributed systems like Hadoop. 
Building on this, subsequent research extended these ideas by integrating adaptive scheduling mechanisms and addressing challenges such as scalability, dynamic resource allocation, and cost awareness, further refining the understanding of optimization strategies in data pipelines.
In this manner, Spark builds on the delayed scheduling mechanism and extends it by introducing in-memory computing through Resilient Distributed Datasets (RDDs), enabling efficient reuse of data across iterative and interactive workloads \citep{zaharia2010spark}.

There is a whole field focusing on optimizing data pipelines within specific domains. 
For example, pipeline optimizations for satellite data were explored by \citep{bacuAdaptiveProcessingEarth2015a}, while others examined optimizations in medical applications \citep{nikolovContainerBasedDataPipelines2023} and smart factory environments \citep{gohConceptualDesignCloudBased2022a}.
These studies typically target the unique requirements of their respective domains, providing valuable insights for those specific contexts. 
However, these findings are often not generalized to other domains, limiting their applicability.

More closely related to our work are studies that provide a wider perspective on data pipeline optimization beyond individual cases. 
For instance, \citet{zhouEScienceBigData2016} examine big data workflows in the cloud and propose a taxonomy of eScience services. 
Their taxonomy encompasses dimensions relevant to our study, such as infrastructure (e.g., cloud), ownership models (e.g., public cloud), and service types (e.g., SaaS or PaaS). 
However, their work primarily focuses on domain-specific cases and includes the development of a heuristic scheduling algorithm aimed at reducing costs while meeting deadline constraints (\autoref{sec:results}).
Building on this foundation, we expand the scope of their taxonomy by incorporating additional optimization dimensions. 
Specifically, we consider a broader range of optimization goals beyond cost reduction under deadline constraints, exploring optimization goals beyond cost-related ones, like makespan, storage, and reactivity. 
This perspective allows us to address gaps in the literature by providing a more comprehensive understanding of user needs and optimization strategies applicable across multiple domains.

\citet{nezamiAnalysisEnergyEfficient2023a} analyze energy efficiency in cloud data centers, with a particular focus on IaaS environments. 
Their work employs scheduling as the primary optimization method to enhance energy efficiency. 
While energy efficiency and sustainability are increasingly critical considerations for SaaS providers, our review of the literature suggests that software-side optimizations primarily target objectives such as minimizing cost and makespan. 
Energy efficiency, though potentially a secondary outcome of these goals, is not typically treated as a primary driver for pipeline optimizations.
In contrast, our work adopts a broader perspective on optimization strategies, which serves as a foundation for evaluating energy efficiency alongside other factors. 
By examining a wider range of optimization approaches beyond scheduling, we aim to provide insights into how diverse strategies impact both traditional metrics, like cost and execution speed, and emerging priorities, such as sustainability.

\citet{barikaOrchestratingBigData2020} provide a comprehensive survey comparing big data programming models and propose a taxonomy to assess and compare big data orchestration challenges effectively. 
Their study addresses a wide range of topics, including security, monitoring, and cloud-related challenges, offering a broad and holistic perspective on data pipeline orchestration.
In contrast, our work focuses specifically on optimization strategies for the execution of data pipelines.
While their research emphasizes orchestration and predefined methods, we aim to summarize and organize a wider array of optimization strategies from the literature, intentionally avoiding limitations to predefined approaches. 
This broader focus allows us to capture the diverse and evolving strategies employed across different contexts, providing a more versatile framework for understanding pipeline optimization.

Systematic literature reviews specifically addressing data pipelines remain relatively sparse. 
\citea{kolajo2019big} provide a notable review, systematically analyzing existing tools and methods for big data stream analysis. 
While our study also considers streaming pipelines, we expand the scope to include batch processing and MapReduce-based data pipelines.
Additionally, our focus diverges in terms of objectives. 
Whereas \citet{kolajo2019big} emphasize tools and technologies, our study centers on understanding optimization goals, contexts, and implementations. 
This distinction enables us to provide a holistic perspective on the strategic considerations driving pipeline optimizations, complementing their tool and technology-focused approach.

\citea{shojaee2024data} conduct another such systematic review on data pipeline frameworks and construct a taxonomy based on their findings. 
In many respects, our study is similar, as both seek to organize knowledge from the literature. 
However, while their work presents its findings as a taxonomy, we structure our contributions as a concept matrix.
Moreover, our study expands the focus in several ways. 
First, we center on data pipelines operating in the cloud, offering a more general perspective compared to their emphasis on serverless frameworks. 
Second, we expand the scope to include a diverse range of optimization goals and contextual factors beyond specific approaches. 
This broader lens enables a more comprehensive understanding of how various optimization strategies apply to cloud-based data pipelines across different use cases.

To the best of our knowledge, there has not been a systematic literature review focusing on application- and domain-independent pipelines from the perspective of cloud service providers. 
Our research fills this gap by providing a structured overview of optimization techniques tailored to general-purpose SaaS and PaaS contexts.  
\section{Research Design} \label{sec:research_design}
\begin{figure}
 \centering
 \includegraphics[width=\linewidth]{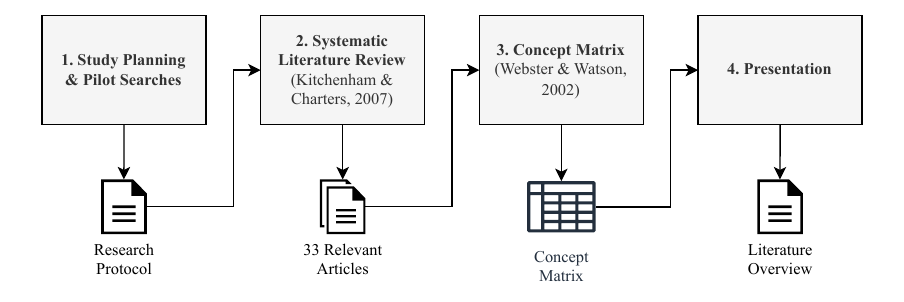}
 \caption{Overview of the complete research process, from initial study design to final presentation.}
 \Description{Step 1: Research Protocol. Step 2: SLR with 33 studies. Step 3: Concept Matrix. Step 4: Presentation}
 \label{fig:slr_process}
\end{figure}

\autoref{fig:slr_process} depicts our process employed in this study.
We conducted a systematic literature review following the guidelines established by \citea{kitchenham_guidelines_2007}.
To ensure a comprehensive understanding of the topic, we familiarized ourselves with existing research and critically assessed the originality and relevance of our study.
These initial insights, along with our research plan, were formally documented and published as a research protocol \citep{jablonski_2024_10725916}.

The final selection of relevant literature comprises 33 studies, whose key insights were systematically captured in a concept matrix.
This concept-centric approach was chosen as it aligns effectively with our research questions, enabling structured synthesis and facilitating the use of findings in future research endeavors.

\subsection{Search Strategy}

To identify relevant literature addressing our research question, we conducted a comprehensive search across multiple academic databases, including the ACM Guide to Computing Literature\footnote{https://libraries.acm.org/digital-library/acm-guide-to-computing-literature}, IEEE Xplore\footnote{https://ieeexplore.ieee.org/Xplore/home.jsp}, Google Scholar\footnote{https://scholar.google.com/}, and Scopus\footnote{https://www.scopus.com/search/form.uri}.
Given the emerging nature of the topic, we did not impose any restrictions on publication years to ensure a broad and inclusive dataset.

Our search strategy was structured around three key thematic pillars, which were combined using logical AND operators to refine the results. Within each pillar, we employed multiple synonymous terms connected by OR operators to maximize coverage and capture variations in terminology. For data pipelines, we decided to include terms related to workflows in the search term as well to capture all potentially relevant papers as data pipelines have also been called workflows in the literature \citep{mesbah_semantic_2017}.

Our search was intentionally broad, designed to capture all papers discussing cloud-based data pipeline systems. We then used the inclusion and exclusion criteria to narrow the focus. The selected pillars are:

\begin{enumerate}
    \item \say{data pipeline}, \say{data processing pipeline}, \say{data workflow}, \say{data processing workflow}, and their plural variants
    \item \say{framework}, \say{architecture}, \say{design}, \say{infrastructure}, and their plural variants
    \item \say{cloud}, and the plural variant \say{clouds}
\end{enumerate}

\subsection{Inclusion and exclusion criteria}

We incorporated content-related criteria into the inclusion criteria and formulated framework conditions as exclusion criteria.
Each paper must fulfill every inclusion and exclusion criteria.
We applied them to each title, abstract, and full text.
The following criteria were used:

\textbf{Inclusion criteria}
\begin{enumerate}
    \item[IC1] The article describes optimization techniques for running multiple data pipelines in the cloud.
    \item[IC2] Data pipelines and their optimizations are presented as majorly application and domain-independent.
    \item[IC3] The article describes software-side optimization techniques.
    \item[IC4] The article includes a qualitative or quantitative evaluation of the optimization technique.
\end{enumerate}

\textbf{Exclusion criteria}
\begin{enumerate}
    \item[EC1] The article is not written in English.
    \item[EC2] The article is not accessible to us in full text.
    \item[EC3] The article has not been published in a peer-reviewed journal, conference proceedings, or workshop.
\end{enumerate}

We designed the selection criteria in a way that they embody several quality criteria.
For example, IC4 only allows the selection of studies with a thorough evaluation, and EC3 ensures all articles passed a proper peer review; both are strong quality criteria for research studies.
EC3, in particular, excludes grey literature such as blog posts or practitioner books that lack academic peer review.

With this approach, we prioritized methodological rigor over the inclusion of, but not peer-reviewed, insights from practitioners.
While this may limit certain perspectives, it ensures a high quality of primary materials and increases the likelihood that our results generalize well to other contexts.
Since this selection by quality was already conducted in this phase, we decided against an additional quality score calculation and filtering based on that.

\subsection{Selection Results} \label{subsec:results:selected-literature}

\begin{figure}
 \centering
 \includegraphics[width=0.9\linewidth]{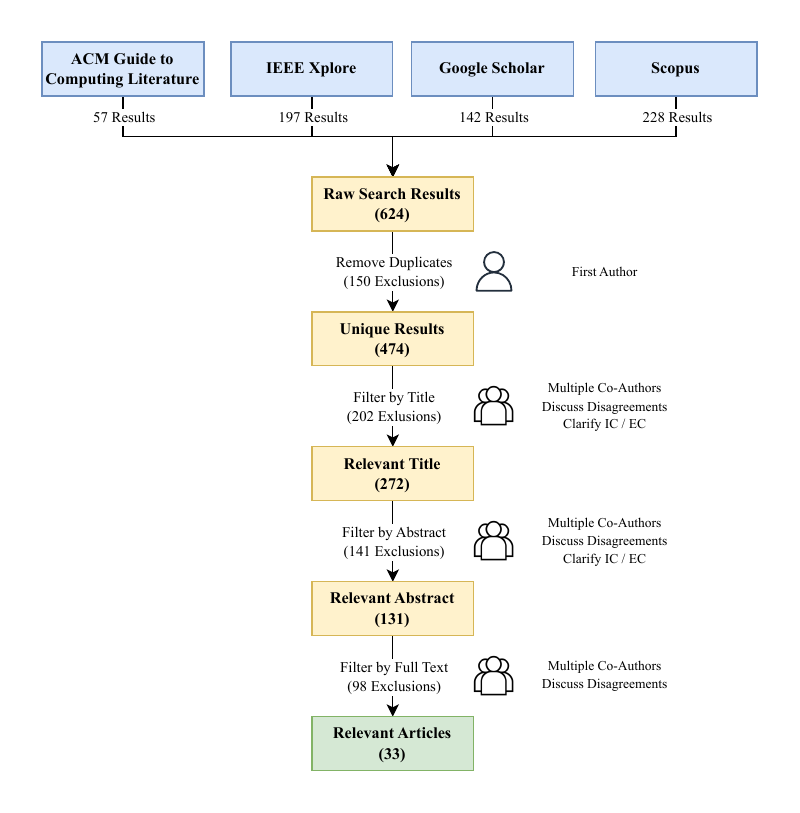}
 \caption{Literature selection process, from the original sources to filtering according to duplicates, title, abstract, and full text by multiple co-authors. The final result is a set of relevant articles.}
 \Description{Articles were selected from ACM Guide to Computing Literature, IEEE Xplore, Google Scholar and Scopus. They were filtered according to duplicates, title, abstract and full text by multiple co-authors, leading to a final set of relevant articles.}
 \label{fig:slr_numbers}
\end{figure}

\begin{figure}
 \centering
 \includegraphics[width=0.9\linewidth]{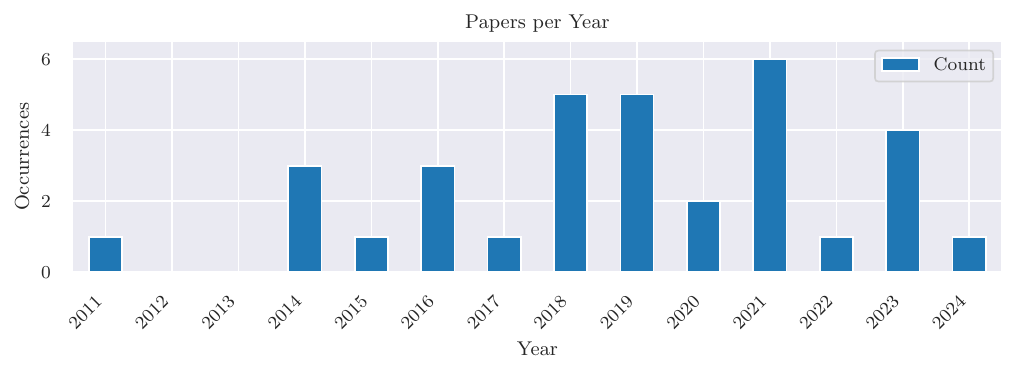}
 \caption{Year distribution of the papers included in the analysis.}
 \Description{The papers analyzed are published between 2011 and 2024 with the majority from 2018 on.}
 \label{fig:publications_per_year}
\end{figure}

We extracted 624 articles, of which 33 were analyzed in detail for this study (\autoref{tab:sources}, \autoref{fig:slr_numbers}).
They were filtered according to duplicates, title, abstract, and full text by multiple coauthors which led to a final set of 33 relevant articles.
If a study could not be clearly rejected by the inclusion and exclusion criteria, it was carried to the next step.
This procedure limits the the risk of studies being inadvertently excluded.
The final set of studies analyzed consists of 16 journal publications and 17 from conferences.
The articles analyzed include studies between 2011 and 2024, with the majority published in 2018 and after (\autoref{fig:publications_per_year}).

\begin{table}[hbt]
    \centering
    \footnotesize
    \caption{Article Sources}
    \label{tab:sources}
    \begin{tabular}{lrrrrrr}
    \toprule
    Source & ACM & IEEE & Scholar & Scopus & $\Sigma$ & Without Duplicates \\
    \midrule
    Exported & 57 & 197 & 142 & 228 & 624 & 474 \\
    Included & 6 & 13 & 17 & 12 & 48 & 33 \\
    \bottomrule
    \end{tabular}
\end{table}

\subsection{Data Analysis}

For analyzing the selected literature, we collected meta-information about the articles, like publication year and publication outlet, to describe the examined population (see Section \ref{subsec:results:selected-literature}).
Further, we used \citeauthor{webster_analyzing_2002}'s concept matrix to analyze the data \citep{webster_analyzing_2002}.
We created a table with a row for each article and columns to capture common patterns (i.e., \textit{concepts}) across the articles related to our research questions.

The columns of the table emerged incrementally during the analysis, each representing a concept under examination to answer our research questions.
Over time, we grouped columns into hierarchical groups that had a cohesive theme (category), e.g., all columns on optimization goals.
In each cell, we marked the contribution of the articles to the concepts, if any.
For data synthesis, we transposed the table, turning this author-centric view into a concept-centric view.
By reading each row of the transposed table, one can read which and how each article contributes to the concept.
Table \ref{tab:matrix_excerpt_goals} shows an excerpt of the concept matrix for the category \textit{optimiation goal}.%
~The complete matrix with all categories can be found in \autoref{tab:matrix}.

\begin{table}[htb]
    \centering
    \footnotesize
    \caption{Excerpt of the concept matrix for the category \textit{optimization goal} (\autoref{subsubsec:optimization_goals}).}
    \begin{tabular}{l|ccccccc}
        \toprule
        Study \textbackslash\ Goal & C & \makecell{C/M} & C-M-Tradeoff & M & M/C & Reactivity & Storage \\
        \midrule
\citepaperalias{291} & X & . & . & . & . & . & X \\
\citepaperalias{579} & . & . & . & X & . & . & . \\
\citepaperalias{322} & . & . & . & X & . & . & . \\
\citepaperalias{49} & . & . & X & X & . & . & . \\
        \bottomrule
    \end{tabular}
    \label{tab:matrix_excerpt_goals}
\end{table}

\subsection{Quality Measures}

We used established quality measures to ensure the validity and objectivity of our review. We started with these activities before the actual review by writing and publishing a review protocol \citep{jablonski_2024_10725916}. Publishing such a research protocol lowers the possibility of researcher bias \citep{kitchenham_guidelines_2007}.

For the literature selection, we applied the suggestions of \citet{perez_systematic_2020} and calculated inter-rater agreement on a subset of studies.
In this manner, the leading researcher conducted the literature selection on a batch of articles.
In parallel, a second and a third researcher conducted the literature selection as well on the same batch of articles.
Afterward, we compared the different selections quantitatively and qualitatively.
For quantitative comparison, we calculated Cohen's kappa score \citep{cohenCoefficientAgreementNominal1960} to measure the pairwise agreement between the researchers.
This procedure strengthens the objectiveness of the inclusion and exclusion criteria \citep{perez_systematic_2020, kitchenham_guidelines_2007}.
A kappa score of 0.8 indicates substantial to almost perfect agreement \citep{landisMeasurementObserverAgreement1977}.
We followed this procedure twice (see Table \ref{tab:study_selection_table}), and reached a kappa score of 1 at the end, indicating very strong agreement among researchers on the selection of literature.
In addition, we also compared the selection of the different researchers qualitatively after each iteration, discussing differences, clarifying ambiguities, and resolving disagreements.
As a result, we refined and extended the inclusion and exclusion criteria to reflect a clear and repeatable selection procedure.

\begin{table}[hbt]
    \centering
    \footnotesize
    \caption{Results of peer-supported study selection. (Criteria with changes only.)}
    \label{tab:study_selection_table}
    \begin{tabular}{lccccl}
    \toprule
    Iteration & $n$ & $k_{12}$ & $k_{23}$ & $k_{13}$ & \\
    \midrule
    Research Protocol (\citep{jablonski_2024_10725916}) & - & - & - & - & \\
    & \multicolumn{5}{l}{
        \parbox{10cm}{
        \textbf{Changed inclusion criteria:}
        \begin{enumerate}
            \item The article describes optimization techniques for running multiple application- and domain-independent data pipelines.
        \end{enumerate}
    }} \\
    Iteration 1 & 63 & 0.86 & 0.53 & 0.69 & \\
    & \multicolumn{5}{l}{
        \parbox{10cm}{
        \textbf{Changed inclusion criteria:}
        \begin{enumerate}
            \item The article describes optimization techniques for running multiple data pipelines in the cloud.
            \item Data pipelines and their optimizations are presented as majorly application and domain-independent.
            \item \textit{Former IC2}
            \item \textit{Former IC3}
        \end{enumerate}
    }} \\
    Iteration 2 & 15 & 1 & 1 & 1 & \\
    \bottomrule
    \end{tabular}
\end{table}

Because the process of developing concepts for the concept matrix is similar to qualitative data analysis, we used inter-coding agreement measurements to evaluate our concept matrix.
To do so, the main analysis via the concept matrix was done by the first author, and at 3 points in time, a snapshot of the concept matrix without the cell values was handed to a second and/or third researcher to replicate the analysis on a subset of the articles (\autoref{tab:inter-coder-agreement}).
Each peer's analysis was compared to the main author's analysis in a similar way to the literature selection.
We computed Cohen's kappa to measure the agreement among pairwise researchers quantitatively.
In the first iteration, we compared researcher 1, 2, and 3 pairwise, resulting in three kappa values $k_{12}$, $k_{13}$, and $k_{23}$.
For the following two iterations, we used the default way of measuring by comparing the outcome of two researchers only.
In addition, we compared the analysis results qualitatively after each iteration to clarify ambiguous interpretations and refine the analysis.
In this manner, we employed investigator triangulation according to \citea{guion2011triangulation} to increase the trustworthiness in our findings.
Our saturation criteria was a kappa value greater than 0.8 and no changes to the concept hierarchy.
We reached that value after 3 iterations with a kappa of 0.849 (\autoref{tab:inter-coder-agreement}), indicating strong agreement among researchers \citep{landisMeasurementObserverAgreement1977}.

\begin{table}[htb]
    \centering
    \footnotesize
    \caption{Inter-coder agreement: Average Cohen's Kappa between the researchers 1, 2, and 3 and number of changes to the concept hierarchy. The scores depict the average over three studies per iteration.}
    \begin{tabular}{lrrrrr}
        \toprule
        Iteration & $k_{12}$ & $k_{13}$ & $k_{23}$ & Concept Hierarchy Changes & Comment Changes \\
        \midrule
        init & - & - & - & 45 & 45 \\
        1 & 0.467 & 0.500 & 0.471 & 3 & 6 \\
        2 & 0.732 & & & 1 & 4 \\
        3 & & 0.849 & & 0 & 1 \\
        \bottomrule
    \end{tabular}
    \label{tab:inter-coder-agreement}
\end{table}

\section{Results} \label{sec:results}
\newcommand{\concept}[1]{\textit{#1}}

We set out to explore how service providers can optimize the cloud-based execution of multiple application- and domain-independent data pipelines (RQ1).
To do so, we designed and executed a systematic literature review (see Section \ref{sec:research_design}).

\autoref{subsec:overview} provides an overview of the results.
\autoref{subsubsec:optimization_goals} presents the optimization goals that justify the need for optimizations.
\autoref{subsec:optimization_context} details the concepts we attribute to the optimization context that define the setting of the pipeline execution.%
\autoref{subsec:optimization_realization} picks up on the concepts of optimization implementations, which describe details on how the goal is reached in a given context.%

\subsection{Overview}
\label{subsec:overview}

\begin{figure}[htb]
 \centering
 \includegraphics[]{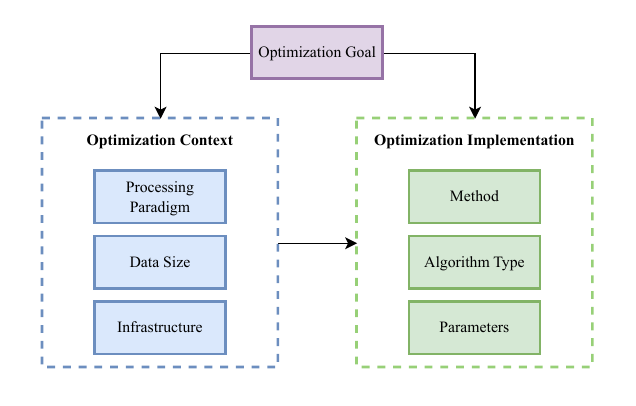}
 \caption{Overview of the concepts discovered through our study.}
 \Description{A overview for optimization approaches in cloud-based data pipeline infrastructures. The optimization context consists of processing paradigm, data size and infrastructure. Together with the optimization goal, this context influences the optimization method(s) that can be chosen. In turn, the optimization method(s) influence algorithm type that are sensible and which of the available parameters are considered. The choice of optimization method(s), optimization parameters and algorithm type taken represent a concrete implementation of an optimization.}
 \label{fig:research_model}
\end{figure}

Figure \ref{fig:research_model} gives an overview of the key concepts drawn from the literature.
It displays higher-order categories, depicted by the boxes (both dashed and filled).
These categories aggregate related concepts of our concept matrix, which are described in detail in the following subsections.
Influencing relationships between these categories are depicted by arrows.

\begin{itemize}
    \item \textbf{Optimization goal}:
    The optimization goal describes the initial trigger of motivation to why an optimization is pursued. 
    Different optimization goals motivate different optimizations, making the goal a major influence on the implementation.
    The goal also influences the context since, for example, specific goals like a high reactivity are only possible with a specific processing paradigm (stream pipelines).

    \item \textbf{Optimization context}:
    Categories in this group describe the overarching condition and characteristics surrounding the execution of data pipelines in a system. 

    \begin{itemize}
        \item \textbf{Data size}:
        The data size refers to the amount of data that is processed per data pipeline.
        It has a major influence on the optimization implementation, especially the optimization method, because different approaches are tailored to different amounts of processed data.
        
        \item \textbf{Processing paradigm}:
        The processing paradigm defines how an application's data is to be processed, for example, in a large batch or in real-time.
        Similarly to the data size, optimization methods each accommodate different processing paradigms and, thus, are influenced by the processing paradigm's manifestation.
        
        \item \textbf{Infrastructure}:
        The infrastructure relates to the environment in which a data pipeline runs; in our context, a specific cloud infrastructure.
        This setting can heavily influence which optimization implementations are feasible and likely to bring benefit.
    \end{itemize}

    \item \textbf{Optimization implementation}:
    Categories in this group describe the components and mechanism through which optimization strategies are implemented.

    \begin{itemize}
        \item \textbf{Method}:
        The optimization method reflects the general idea behind a concrete optimization, the mechanism on which the optimization is built.
        Most studies employed a combination of several methods per optimization instead of relying on just one.
        
        \item \textbf{Algorithm type}:
        The algorithm type describes how the corresponding optimization is calculated.
        We observed that the algorithm is influenced by the requirements of the optimization method most but, in general, is often freely choosable.
        
        \item \textbf{Parameters}:
        The parameters describe which parameters are considered in the optimization.
        Some parameters might be used for additional functionality, such as monitoring.
        However, since this study is tailored to the execution of data pipelines, we only included parameters that are used to improve execution.
    \end{itemize}
\end{itemize}

The following subsections detail the concepts of the optimization goals (\autoref{subsubsec:optimization_goals}), contexts (\autoref{subsec:optimization_context}) and the implementations (\autoref{subsec:optimization_realization}).

\subsection{Optimization Goals} 
\label{subsubsec:optimization_goals}

We found seven major optimization goals in the examined literature to answer (\ref{rq:goals}).
\autoref{tab:occurances_optimization_goal} gives an overview of the optimization goals and their frequency of occurrence, highlighting that the makespan and cost-oriented goals are the most common.
The sum of optimization goals found exceeds the number of examined studies since several articles present multiple separate optimization goals, particularly when presenting and comparing multiple optimizations.

\begin{table}[ht]
    \centering
    \footnotesize
    \caption{List of optimization goals}
    \label{tab:occurances_optimization_goal}
    \begin{tabularx}{\textwidth}{lrX}
        \toprule
        Goal & Count & Studies \\
        \midrule
Cost & 4 & \citepaperalias{604}, \citepaperalias{424}, \citepaperalias{291}, \citepaperalias{224} \\
Cost Under Makespan Constraint & 6 & \citepaperalias{19}, \citepaperalias{373}, \citepaperalias{515}, \citepaperalias{450}, \citepaperalias{511}, \citepaperalias{326} \\
Cost-Makespan-Tradeoff & 9 & \citepaperalias{182}, \citepaperalias{119}, \citepaperalias{593}, \citepaperalias{133}, \citepaperalias{49}, \citepaperalias{511}, \citepaperalias{334}, \citepaperalias{126}, \citepaperalias{431} \\
Makespan & 13 & \citepaperalias{449}, \citepaperalias{194}, \citepaperalias{579}, \citepaperalias{322}, \citepaperalias{49}, \citepaperalias{379}, \citepaperalias{35}, \citepaperalias{474}, \citepaperalias{83}, \citepaperalias{623}, \citepaperalias{487}, \citepaperalias{554}, \citepaperalias{229} \\
Makespan Under Cost Constraint & 4 & \citepaperalias{511}, \citepaperalias{514}, \citepaperalias{442}, \citepaperalias{426} \\
Reactivity & 1 & \citepaperalias{487} \\
Storage & 2 & \citepaperalias{424}, \citepaperalias{291} \\
        \bottomrule
    \end{tabularx}
\end{table}

\paragraph{Cost}
The optimization goal cost refers to minimizing the monetary expenses of data pipelines.
Studies optimizing for cost typically focus on the execution cost but, in general, may also consider transfer or storage costs.
For example, \citea{291_ikken_2018} break costs down into data transfer costs in or between data centers, data movement costs of partial intermediate data, and storage costs for data stored in one data center.
\citea{424_fanOptimizingDataRegeneration2024} take a simpler approach by setting aside transfer costs and time and focusing on the price of storage and computing services.

\paragraph{Makespan}
Makespan refers to the time it takes from the initialization of a pipeline until its end.
In addition to the pure duration of the execution of a pipeline, \citea{19_ahmadDynamicVMProvisioning2021} also include potential waiting time before the start in the form of acquisition delay.
This delay typically exists because VMs need to be provisioned or the scheduling decisions need to be calculated.
For stream pipelines that are potentially indefinite, makespan refers to the duration of a data tuple or batch from the start to the end of the stream.

\paragraph{Cost-Makespan-Tradeoff}

Finding the minimum cost and the minimum makespan of a data pipeline is often a contradiction \citep{119_bousselmi_bi_objective}.
Nevertheless, optimizing for both is a common use case, with a tradeoff required at some end.
The sweet spot for such a tradeoff is context-dependent, so it might be biased toward one end instead of striving for a perfect balance.
Optimizations do not necessarily need to return exactly one prediction. For example, \citea{182_aseman-manzarCostAwareResourceRecommendation2023} generates a set of different makespan-cost predictions to choose from.

\paragraph{Cost Under Makespan Constraint}

The goal \textit{Cost Under Makespan Constraint} is similar to the goal of minimizing costs, with the difference that a certain time must not be exceeded.
While the limit could be a duration with or without an initial delay until the pipeline starts executing, we only observed pipelines that do include the initial delay in their calculations.
Furthermore, the limit can be a duration (i.e. relative time) or a fixed point in time (i.e. absolute time).
For example, \citea{511_kllapiScheduleOptimizationData2011} present a system where the system's client can decide on the exact deadline.

\paragraph{Makespan Under Cost Constraint}

The goal \textit{Makespan Under Cost Contraint} is the reverse of the goal \textit{Cost Under Makespan Constraint}.
It is similar to a makespan optimization but with an upper limit for the cost.
The execution time of a data pipeline should be as low as possible without exceeding a fixed maximum budget.
For example, \citea{442_shuPerformanceOptimizationHadoop2017} focus on MapReduce workflow mappings in Hadoop/YARN. \citea{426_wuOptimizingPerformanceBig2016} propose a workflow scheduling solution for IaaS providers in multi-cloud environments, both optimizing makespan under cost constraint.

\paragraph{Reactivity}

In stream pipelines, data usually comes in as a continuous flow of data.
Optimizing for \textit{Reactivity} results in minimizing the delay of data changes and typically leads to a good (near) real-time performance.
The difference between \textit{Reactivity} and \textit{Makespan} is that \textit{Reactivity} optimizes the wait until the pipeline begins and the latter minimizes the execution time of the pipeline from start to finish.
For example, \citea{487_tranReducingTailLatencies2018} contribute an approach to minimize tail latency (the 99\% percentile of latencies) as a proxy for reactivity by proposing their programming model Dynamo of duplication in stream processing frameworks.

\paragraph{Storage}

Every data pipeline processes and stores data at various stages—initial ingestion, intermediate processing, and final output, all of which can be subject to optimizations. 
\citea{291_ikken_2018} focus specifically on optimizing intermediate \textit{Storage} by proposing a greedy heuristic algorithm to deal with the placement of potentially unsplittable data between multiple data centers.
In contrast, \citea{424_fanOptimizingDataRegeneration2024} propose an approach to optimize final long-term storage by identifying and deleting unused data. If the deleted data is subsequently required, it can be reconstructed using the preserved data pipeline.

\subsection{Optimization Context} 
\label{subsec:optimization_context}

The optimization context describes the environment considered when designing and deciding on an optimization implementation for data pipelines.
We discovered three categories: infrastructure, data size, and processing paradigm.

\subsubsection{Infrastructure}
\label{subsubsec:infrastructure}

\begin{figure}[ht]
  \centering
  \includegraphics[width=0.8\linewidth]{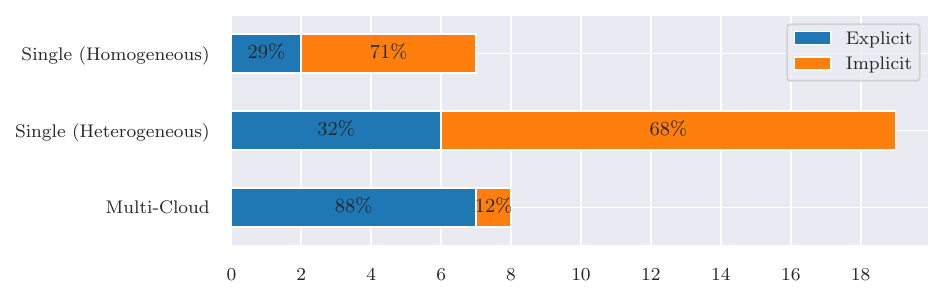}
  \caption{Distribution of different infrastructures types}
  \label{fig:occurences_infrastructure}
  \Description{Bar chart with three bars. Single-Cloud (Heterogeneous) is by far the most used concept before Multi-Cloud and Single-Cloud (Homogeneous)}
\end{figure}

The infrastructure concept, as part of the optimization context, refers to the environment where the data pipeline is operated in the cloud.
We found that 26 (79\%) of optimization studies rely on a single-cloud setup within one data center while 8 (24\%) consider a multi-cloud setup using multiple data centers.
Our analysis revealed that most studies focus exclusively on one type of infrastructure, with \citet{579_kacsukFlowbsterCloudOrientedWorkflow2018} standing out as the only work providing an algorithm applicable to both single-cloud and multi-cloud scenarios.
Further, we distinguished whether the cloud setup considered different types of VMs (heterogeneous cloud) or only a single one (homogeneous cloud). 
However, this information was only explicitly mentioned by less than half of the studies (see \autoref{fig:occurences_infrastructure}).
We inferred the infrastructure setup for the remaining articles, typically by drawing on their evaluation details.

\paragraph{Homogeneous Single Cloud Infrastructure}
In a homogeneous single-cloud deployment, only one type of VM is used even if multiple different VM types are available.
A VM type here refers to one configuration of a computing resource and typically contains, amongst others, a fixed RAM, CPU, storage, and network capacity.
This setup was rarely described as explicitly as in \citet{424_fanOptimizingDataRegeneration2024}, where it is explicitly referred to a \say{single homogeneous cloud environment}.
Instead, we deduced this information by examining the test setups provided in the studies.

\paragraph{Heterogeneous Single Cloud Infrastructure}
A common configuration in single-cloud deployments is a heterogeneous cloud, where a single data center provides a diverse range of VM types.
Even within a single data center, optimization algorithms can leverage this diversity to select the most suitable VMs.
\citea{49_kashlevSystemArchitectureRunning2014} work with the common assumption that these VMs share similar capabilities.
In contrast, In contrast, \citea{442_shuPerformanceOptimizationHadoop2017} exploit the distinct specializations of certain VMs, such as storage-centric designs or GPU support, to further enhance performance.

\paragraph{Multi-Cloud Infrastructure}
Multi-cloud infrastructures stand in contrast to single-cloud deployments, as they span multiple data centers that are typically geographically distributed.
In the context of this review, studies focusing on multi-cloud infrastructures consistently employ heterogeneous VM types.
Optimizations targeting such infrastructures may need to account for increased challenges between geographically distributed infrastructures, for example as done by \citea{229_zhangDataintensiveWorkflowScheduling2023} and \citea{379_mehranMatchingbasedSchedulingAsynchronous2022}.
\citea{579_kacsukFlowbsterCloudOrientedWorkflow2018} explore a multi-cloud setup using a hybrid cloud configuration to enable a balance of performance and cost considerations.

\subsubsection{Data Size}
\label{subsubsec:data-size}

Another contextual factor for data pipeline optimizations is the quantity of data processed per workflow.
To define whether or not a study presented optimizations for big data use-cases, we first tried to extract the target data size of the described solutions directly from the manuscripts to present the work in the context the authors intended. This was possible because most studies explicitly reported on their assumptions about processed data size.

If we could not find an explicit mention of data size, we inferred it from the workflows and data used in the evaluation, following the well-known \say{3 Vs} definition by volume, velocity, and variety by \citea{laney01controlling3v} and the characteristic of whether it can be processed on a single machine or not \citep{scottBayesBigData2016}.

\begin{figure}[ht]
  \centering
  \includegraphics[width=0.66\linewidth]{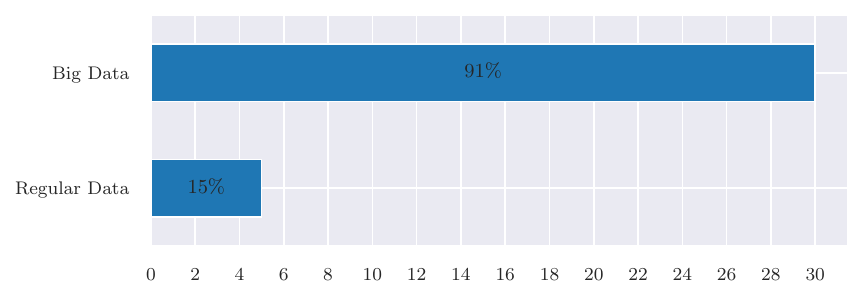}
  \caption{Distribution of data sizes}
  \Description{Bar chart with two bars which shows 30 studies about big data and 5 studies about regular data.}
  \label{fig:occurences_data_sizes}
\end{figure}

\paragraph{Big Data}
Big data is the most common use case in the literature studied and covered by 30 of the 33 studies.
Several known big data workflows are frequently employed across multiple studies to evaluate optimization strategies. Notably, the Montage \citep{jacob_2009_montage}, LIGO \citep{deeplman_2002_LIGO}, and Cybershake \citep{deelman_2006_cybershake} workflows were all utilized in studies \citet{83_nascimentoIncrementalReinforcementLearning2021}, \citet{511_kllapiScheduleOptimizationData2011}, and \citet{326_mousavimojabICATSSchedulingBig2019}. These workflows appear as standard benchmarks in the field, with Montage being featured in 10 studies, LIGO in 8, and Cybershake in 5.

\paragraph{Regular data}
Regular data covers all data that are not large enough to fall into the big data category. Solely regular data is covered by three of the 33 studies, whereas two more studies cover big data in addition to regular data.
These optimizations do not have to cope with large data volumes but rather focus on other factors.
For instance, \citea{379_mehranMatchingbasedSchedulingAsynchronous2022} applied their optimization approach to two experiments, each requiring less than 2GB of working memory and storage. Similarly, \citea{474_muller2014pydron} evaluated their optimization within a cloud environment handling approximately 3.8GB of data.
Notably, both cases involved machine learning components within their data pipelines.
While the volume of data may appear modest, the computational requirements of these pipelines remain non-trivial due to the complexity of the calculations involved.

\subsubsection{Processing paradigm}
\label{subsubsec:processing-type}

\begin{figure}[ht]
  \centering
  \includegraphics[width=0.66\linewidth]{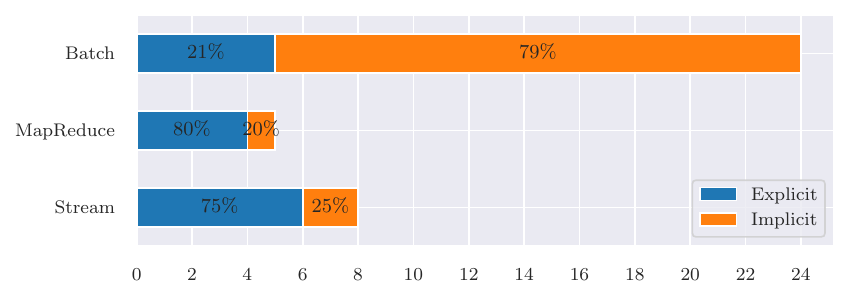}
  \caption{Distribution of processing paradigm}
  \Description{A bar chart showing the 3 bars: Batch (24), MapReduce (5), and Stream (8).}
  \label{fig:occurences_processing_types}
\end{figure}

We found three different processing paradigms for data pipelines (see \autoref{fig:occurences_processing_types}): batch, stream, and map-reduce-based processing.
24 studies (73\%) report optimizations on batch processing, eight studies (24\%) focus on stream processing, and five studies (15\%) present map-reduce-based optimizations. 
Two studies (\citet{322_kamburugamuveHPTMTOperatorBasedArchitecture2021} and \citet{511_kllapiScheduleOptimizationData2011}) support both stream and batch processing optimizations.
We found that almost two-thirds (64\%) of the studies do not mention their processing paradigm.
Similarly to the handling of infrastructure characteristics (\autoref{subsubsec:infrastructure}), we inferred the processing paradigm in such cases.
This was particularly common in batch processing studies (79\%).

\paragraph{Batch Processing}
In batch pipelines, the whole dataset must be present before starting the execution.
A common implementation is found in ETL pipelines. %
Although the full data are processed at once, this does not imply that the whole data set must be processed atomically.
For instance, \citea{83_nascimentoIncrementalReinforcementLearning2021} partition the input data into several smaller segments to facilitate execution.

\paragraph{Map-Reduce-based Processing}
A more restrictive variation of batch processing is MapReduce.
Here, the execution is highly parallelized into map and reduce jobs across a computation cluster and is therefore well-suited for processing large amounts of data \citep{deanMapReduceSimplifiedData2008}.
For instance, \citea{449_chenPISCESOptimizingMultiJob2019} introduce an approach to optimize MapReduce jobs by implementing efficient parallelization techniques for interdependent tasks.
\citea{554_yeStorageAwareTaskScheduling2018} take a different approach and optimize MapReduce jobs by improving data placement and data transfer.

\paragraph{Stream Processing}
Stream processing pipelines do not require the full dataset present before starting execution.
Instead, data is expected to arrive incrementally as it is generated.
In other words, there is no need to feed data into a pipeline immediately.
For example, \citea{450_esteves_2014a} require incoming data to be of a certain level of \say{impact} before being passed to subsequent stages. 

\subsection{Optimization Implementation} 
\label{subsec:optimization_realization}

The optimization implementation describes the details about how a data pipeline optimization reaches the optimization goal.
We discovered three major categories, which we detail one after another in this subsection: method, parameters, and algorithm type.

\subsubsection{Method} 
\label{subsubsec:realization_opti_method}

Optimization methods are fundamental techniques for optimizing data pipelines.
The most prominent method is scheduling, with 27 studies (82\%) using it for at least part of the optimization (see \autoref{tab:occurances_optimization_method}).
We observed that studies often rely on a combination of optimization methods.
Instead of interpreting how much impact each technique had, we assigned multiple techniques to such studies.

\begin{table}[hbt]
    \centering
    \footnotesize
    \caption{List of methods used for optimizations}
    \label{tab:occurances_optimization_method}
    \begin{tabularx}{\textwidth}{lrX}
        \toprule
        Method & Count & Studies \\
        \midrule
Caching & 1 & \citepaperalias{431} \\
Data Placement & 7 & \citepaperalias{449}, \citepaperalias{604}, \citepaperalias{291}, \citepaperalias{224}, \citepaperalias{35}, \citepaperalias{431}, \citepaperalias{554} \\
Data Regeneration & 1 & \citepaperalias{424} \\
Data Transfer & 8 & \citepaperalias{604}, \citepaperalias{49}, \citepaperalias{511}, \citepaperalias{224}, \citepaperalias{431}, \citepaperalias{426}, \citepaperalias{554}, \citepaperalias{229} \\
Execution Overlapping & 2 & \citepaperalias{449}, \citepaperalias{579} \\
Intra-Task Parallelism & 9 & \citepaperalias{194}, \citepaperalias{291}, \citepaperalias{579}, \citepaperalias{322}, \citepaperalias{35}, \citepaperalias{474}, \citepaperalias{83}, \citepaperalias{442}, \citepaperalias{554} \\
Model Transformation & 6 & \citepaperalias{19}, \citepaperalias{604}, \citepaperalias{450}, \citepaperalias{579}, \citepaperalias{474}, \citepaperalias{229} \\
Provisioning & 13 & \citepaperalias{19}, \citepaperalias{182}, \citepaperalias{373}, \citepaperalias{133}, \citepaperalias{515}, \citepaperalias{450}, \citepaperalias{579}, \citepaperalias{224}, \citepaperalias{35}, \citepaperalias{514}, \citepaperalias{326}, \citepaperalias{126}, \citepaperalias{426} \\
Redundancy & 2 & \citepaperalias{133}, \citepaperalias{487} \\
Scheduling & 27 & \citepaperalias{19}, \citepaperalias{182}, \citepaperalias{373}, \citepaperalias{119}, \citepaperalias{593}, \citepaperalias{133}, \citepaperalias{449}, \citepaperalias{604}, \citepaperalias{194}, \citepaperalias{515}, \citepaperalias{450}, \citepaperalias{579}, \citepaperalias{49}, \citepaperalias{511}, \citepaperalias{224}, \citepaperalias{379}, \citepaperalias{35}, \citepaperalias{514}, \citepaperalias{326}, \citepaperalias{474}, \citepaperalias{83}, \citepaperalias{334}, \citepaperalias{623}, \citepaperalias{442}, \citepaperalias{426}, \citepaperalias{554}, \citepaperalias{229} \\
        \bottomrule
    \end{tabularx}
\end{table}

\paragraph{Scheduling}
Scheduling defines the location (i.e., the VM and data center) and timing (i.e., the order of execution of subsequent stages) of the execution of individual parts of a pipeline.
With scheduling, we also refer to mapping techniques since the differentiation in literature was only marginal and very occasional.
For example, \citea{133_chenBigDataProcessing2020} delay the scheduling of specific workflow tasks to safe cost without sacrificing execution speed.
\citea{83_nascimentoIncrementalReinforcementLearning2021} avoid relying on a predefined scheduling model and instead use a reinforcement learning approach that adapts based on prior executions of a data pipeline.

\paragraph{Provisioning}
While scheduling determines which VMs are used, provisioning determines which VMs are available.
Provisioning is defined as the in-advance or in-vivo renting and releasing of resources from a cloud provider.
For example, \citea{19_ahmadDynamicVMProvisioning2021} propose an algorithm to detect idle VMs to de-provision, aiming to reduce operational costs.

\paragraph{Intra-Task Parallelism}
Introducing parallelism can improve the performance of pipelines.
Intra-task parallelism is an approach to improve pipeline performance by introducing parallelism.
A single task is broken into smaller sub-tasks so it can be processed on multiple machines simultaneously.
\citea{83_nascimentoIncrementalReinforcementLearning2021} use exactly this approach to lower the makespan of big data batch pipelines.
Pipelines that are build on MapReduce generally include a form of intra-task parallelism already \citep{554_yeStorageAwareTaskScheduling2018}.

\begin{figure}
    \centering
    \begin{subfigure}{.4\textwidth}
      \centering
      \includegraphics[width=1\linewidth]{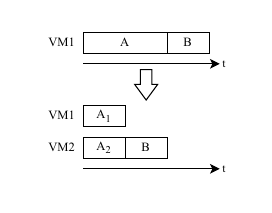}
      \caption{Intra-Task Parallelism}
      \Description{One Task is split up and distributed over 2 VMs now}
      \label{fig:example_intra_task_parallelism}
    \end{subfigure}%
    \begin{subfigure}{.4\textwidth}
      \centering
      \includegraphics[width=1\linewidth]{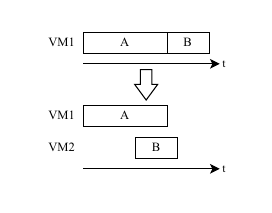}
      \caption{Execution Overlapping}
      \Description{Tasks A and B are not executed sequentially but in parallel on different VMs.}
      \label{fig:example_execution_overlapping}
    \end{subfigure}
    \caption{A comparison of the two types of parallelization found.}
    \label{fig:examples_parallelism}
\end{figure}

\paragraph{Execution Overlapping}
Another approach to introduce parallelism is execution overlapping.
Instead of waiting for previous steps to finish, subsequent stages can start as soon as some data is available (see \autoref{fig:examples_parallelism} for a visual comparison to intra-task parallelism).
\citea{449_chenPISCESOptimizingMultiJob2019} use that to reduce makespan of their MapReduce pipelines.
\citea{579_kacsukFlowbsterCloudOrientedWorkflow2018}) apply this approach to stream pipelines.

\paragraph{Caching}
Caching utilizes temporarily storing intermediate results of pipelines for later use.
For example, \citea{431_tudoranOverFlowMultiSiteAware2016} use this technique to improve the makespan of geo-distributed data pipelines. 

\paragraph{Data Regeneration}
The opposite of caching is data regeneration.
Here, results are discarded and recreated only when needed.
\citea{424_fanOptimizingDataRegeneration2024} use this method to optimize its storage solution while also reducing pipeline cost.

\paragraph{Redundancy}
Partial pipelines are also executed multiple times when optimizing via redundancy.
Segments are executed in parallel, which reduces the time required for certain parts of a data pipeline.
In the literature reviewed, this optimization method is only used in stream pipelines for reactivity and makespan-related improvements.
For instance, \citea{487_tranReducingTailLatencies2018} employ this method to address timing errors effectively, which may arise due to pauses from garbage collection or sudden increases in network latency.
This enables them to increase the reactivity of their stream pipelines.

\paragraph{Data placement}
Any data produced must be stored somewhere, even if it is only in temporary.
Data placement refers to optimizing where to place data, including the data sources, intermediate data, and data sinks.
This data locality problem is not only relevant for multi-cloud setups (e.g., \citet{291_ikken_2018} and \citet{431_tudoranOverFlowMultiSiteAware2016}) but also occurs in single-cloud infrastructures.
For example, \citea{554_yeStorageAwareTaskScheduling2018} utilize knowledge about data locality to lower the execution time of MapReduce pipelines.
We also aggregated optimizations for storing data in general in this optimization, such as using specialized file formats.

\paragraph{Data transfer}
While the data placement method focuses on where and how the data is stored, the data transfer optimization method stands for optimizing the way the data gets to its destination.
This includes the transfer to a pipeline cloud service, the transfer of intermediate data, and also the transfer to a final sink destination.
For example, \citea{426_wuOptimizingPerformanceBig2016} search for optimal bandwidth between clouds to minimize makespan of a pipeline execution a given budget constraint.
This approach is not limited to multi-cloud setups, as \citea{511_kllapiScheduleOptimizationData2011} consider data transfer in their single-cloud optimization approach.

\paragraph{Model transformation}
Model transformation stands out from the previous optimization methods because it analyzes and modifies the pipeline model without changing the semantics or results of a data pipeline.
For example, \citea{604_chenTransformationBasedStreamingWorkflow2019} join or split steps to efficiently limit costs.

\subsubsection{Parameters}

Optimizations utilize different parameters to approximate a global or local optimum related to their optimization goal.
\autoref{tab:occurances_parameters_cosidered} gives an overview of the parameters we found to be considered in the literature.
Please note that we only recorded parameters that contribute to the optimization itself. This excludes, for example, parameters that are just used for monitoring or logging.
In the following, we detail the most common parameters that are considered in optimizations.

\begin{table}[hbt]
    \centering
    \footnotesize
    \caption{List of parameters used for optimizations}
    \label{tab:occurances_parameters_cosidered}
    \begin{tabularx}{\textwidth}{lrX}
        \toprule
        Parameter & Count & Studies \\
        \midrule
Cloud / Data Transfer Price & 8 & \citepaperalias{373}, \citepaperalias{119}, \citepaperalias{604}, \citepaperalias{515}, \citepaperalias{291}, \citepaperalias{326}, \citepaperalias{431}, \citepaperalias{426} \\
Cloud / Data Transfer Rate & 18 & \citepaperalias{19}, \citepaperalias{373}, \citepaperalias{119}, \citepaperalias{593}, \citepaperalias{133}, \citepaperalias{515}, \citepaperalias{291}, \citepaperalias{511}, \citepaperalias{224}, \citepaperalias{379}, \citepaperalias{514}, \citepaperalias{326}, \citepaperalias{83}, \citepaperalias{623}, \citepaperalias{431}, \citepaperalias{426}, \citepaperalias{554}, \citepaperalias{229} \\
Cloud / VM-Types & 17 & \citepaperalias{19}, \citepaperalias{182}, \citepaperalias{373}, \citepaperalias{593}, \citepaperalias{133}, \citepaperalias{515}, \citepaperalias{450}, \citepaperalias{579}, \citepaperalias{49}, \citepaperalias{514}, \citepaperalias{326}, \citepaperalias{83}, \citepaperalias{623}, \citepaperalias{442}, \citepaperalias{126}, \citepaperalias{426}, \citepaperalias{554} \\
Task / Data Size & 14 & \citepaperalias{373}, \citepaperalias{119}, \citepaperalias{133}, \citepaperalias{449}, \citepaperalias{515}, \citepaperalias{291}, \citepaperalias{224}, \citepaperalias{379}, \citepaperalias{35}, \citepaperalias{514}, \citepaperalias{326}, \citepaperalias{431}, \citepaperalias{426}, \citepaperalias{554} \\
Task / Dependencies & 27 & \citepaperalias{19}, \citepaperalias{182}, \citepaperalias{373}, \citepaperalias{119}, \citepaperalias{593}, \citepaperalias{133}, \citepaperalias{449}, \citepaperalias{604}, \citepaperalias{194}, \citepaperalias{515}, \citepaperalias{424}, \citepaperalias{291}, \citepaperalias{579}, \citepaperalias{511}, \citepaperalias{224}, \citepaperalias{379}, \citepaperalias{35}, \citepaperalias{514}, \citepaperalias{326}, \citepaperalias{83}, \citepaperalias{623}, \citepaperalias{442}, \citepaperalias{487}, \citepaperalias{431}, \citepaperalias{426}, \citepaperalias{554}, \citepaperalias{229} \\
Task / RAM & 5 & \citepaperalias{450}, \citepaperalias{511}, \citepaperalias{379}, \citepaperalias{442}, \citepaperalias{554} \\
Task / CPU & 15 & \citepaperalias{373}, \citepaperalias{119}, \citepaperalias{593}, \citepaperalias{449}, \citepaperalias{515}, \citepaperalias{450}, \citepaperalias{511}, \citepaperalias{224}, \citepaperalias{379}, \citepaperalias{514}, \citepaperalias{326}, \citepaperalias{83}, \citepaperalias{442}, \citepaperalias{126}, \citepaperalias{554} \\
Past Performance & 7 & \citepaperalias{182}, \citepaperalias{373}, \citepaperalias{450}, \citepaperalias{224}, \citepaperalias{514}, \citepaperalias{83}, \citepaperalias{431} \\
Source Code & 4 & \citepaperalias{194}, \citepaperalias{579}, \citepaperalias{322}, \citepaperalias{474} \\
VM / Billing Cycle & 9 & \citepaperalias{19}, \citepaperalias{373}, \citepaperalias{593}, \citepaperalias{133}, \citepaperalias{450}, \citepaperalias{511}, \citepaperalias{224}, \citepaperalias{442}, \citepaperalias{426} \\
VM / Boot Delay & 4 & \citepaperalias{19}, \citepaperalias{373}, \citepaperalias{334}, \citepaperalias{126} \\
VM / CPU & 15 & \citepaperalias{373}, \citepaperalias{593}, \citepaperalias{604}, \citepaperalias{515}, \citepaperalias{424}, \citepaperalias{511}, \citepaperalias{379}, \citepaperalias{326}, \citepaperalias{83}, \citepaperalias{334}, \citepaperalias{442}, \citepaperalias{431}, \citepaperalias{426}, \citepaperalias{554}, \citepaperalias{229} \\
VM / Price & 21 & \citepaperalias{19}, \citepaperalias{182}, \citepaperalias{373}, \citepaperalias{119}, \citepaperalias{593}, \citepaperalias{133}, \citepaperalias{604}, \citepaperalias{515}, \citepaperalias{450}, \citepaperalias{424}, \citepaperalias{291}, \citepaperalias{511}, \citepaperalias{224}, \citepaperalias{514}, \citepaperalias{326}, \citepaperalias{83}, \citepaperalias{334}, \citepaperalias{442}, \citepaperalias{126}, \citepaperalias{431}, \citepaperalias{426} \\
VM / RAM & 7 & \citepaperalias{511}, \citepaperalias{379}, \citepaperalias{334}, \citepaperalias{442}, \citepaperalias{431}, \citepaperalias{426}, \citepaperalias{554} \\
VM / Storage & 8 & \citepaperalias{424}, \citepaperalias{291}, \citepaperalias{379}, \citepaperalias{83}, \citepaperalias{334}, \citepaperalias{431}, \citepaperalias{426}, \citepaperalias{554} \\
        \bottomrule
    \end{tabularx}
\end{table}

\paragraph{Cloud}

The following parameters refer to characteristics or resources regarding the cloud infrastructure in which the data pipeline is embedded. 25 of 33 studies considered cloud-related parameters.

The \concept{data transfer price} represents the monetary cost associated with moving data within or across cloud environments.
Specifically, this includes both inter-cloud transfer, where data is moved between distinct cloud providers or regions, and intra-cloud transfer, which refers to data movement within a single cloud provider's infrastructure.
For example, \citea{426_wuOptimizingPerformanceBig2016} focus on optimizing inter-cloud data transfer to minimize costs, while \citea{326_mousavimojabICATSSchedulingBig2019} explore strategies for reducing intra-cloud transfer expenses.
Time-related information about data transfer is captured via the \concept{data transfer rate}.
This information is typically expressed in bandwidth. %
Similarly, \citea{83_nascimentoIncrementalReinforcementLearning2021} incorporate latency considerations into their approach.
Additionally, live data about traffic load plays a role in optimization strategies, as explored by \citea{119_bousselmi_bi_objective}.

A cloud center typically offers multiple \concept{VM types}, categorized based on resource specialization. For example, \citet{442_shuPerformanceOptimizationHadoop2017} analyze different VM configurations available in Amazon EC2, distinguishing between general-purpose, computation-optimized, and storage-optimized instances and instances with GPU. 

\paragraph{Individual Tasks}

The following parameters are specific to individual steps of a data pipeline.
These parameters refer to the characteristics and resource demands of parts of a data pipeline.
29 of 33 studies considered such parameters regarding individual tasks.

\concept{Dependencies} show which steps are required for another step to start.
This is often done by analyzing the DAG
Task dependencies are the most used parameters, mentioned in 27 out of the 33 studies analyzed.
For instance, \citea{579_kacsukFlowbsterCloudOrientedWorkflow2018} leverage the DAG structure to identify and merge workflow tasks to simplify the pipeline execution.
\citea{291_ikken_2018} use the task dependency information to strategically place intermediate data, leading to less data movement and cost.

Tasks are often characterized by an (estimated) \concept{data size} that needs to be processed, typically expressed in bytes.
Optimizations often rely on this data to determine the appropriate machine size to be acquired from a cloud provider or to predict and predict task runtime for a given machine.
One approach to get information about the data size, as adopted by \citea{326_mousavimojabICATSSchedulingBig2019}, integrates both input and output data sizes within the workflow definition.
An alternative strategy, presented by \citea{449_chenPISCESOptimizingMultiJob2019}, uses the input data size to predict the output data size and processing speed of each step.

Information regarding the (estimated) \concept{CPU} requirement frequently influences optimization decisions.
Typically, this requirement is quantified in terms of \say{millions of instructions} needed per task, as illustrated by \citea{379_mehranMatchingbasedSchedulingAsynchronous2022} and \citea{450_esteves_2014a}.

Similarly, the (estimated) \concept{RAM} requirements are also essential for many optimizations.
This information is usually expressed as the amount of memory in bytes. For example, \citet{450_esteves_2014a} demand the amount of memory required for a task to be given in MB.

\paragraph{VM} 
\label{subsub:vm}

This collection of parameters is related to a compute node. 
We used the name VM for simplicity and because VMs are the most prominent version of compute nodes found in the literature studied.
However, we also refer to other cloud resources with these parameters, such has physical machines.

The \concept{price} for a VM is a critical factor in cost-related optimizations.
Typically, cloud providers offer several compute instances with individual pricing per instance and region.
This pricing information can be leveraged to optimize for cost or incorporate price-related constraints.
For example, \citea{182_aseman-manzarCostAwareResourceRecommendation2023} use a cloud provider's cost function to derive two different optimization strategies: one targeting minimal cost and the other aiming for minimal makespan under a given budget limit.
Cloud providers often charge VM usage in integer time units but not all studies include this \concept{billing cycle} in their optimizations.
For example, \citea{133_chenBigDataProcessing2020} and \citea{224_makhlouf2019data} work with a 60 minute lease time of a VMs, with fractions of an hour being rounded up. 
\citeauthor{450_esteves_2014a} describe a Workflow-as-a-Service (WaaS) product where its users pay primarily only for the time of execution, but with a 10\% surcharge to compensate for the cost of idle VMs that are acquired and payed for by the WaaS, but do not execute pipelines \citep{450_esteves_2014a}.

VMs are typically not immediately available upon request, resulting in a \concept{boot delay}.
This delay is taken into account when defining a run schedule, as exemplified by \citea{19_ahmadDynamicVMProvisioning2021} and \citea{373_changxin_lpod}.
\citea{334_quinnImplicationsAlternativeServerless2021} demonstrate an alternative approach to mitigating this cold-start delay by pre-provisioning resources so that they are immediately available when the load increases.

The execution time is greatly influenced by the speed of the processor (\concept{CPU}). 
For example, \citea{593_bugingoDecompositionBasedMultiobjective2021} consider the CPU frequency of VMs as a major part of the pay-as-you-go charging model in the cloud and, thus, adjust the CPU frequency of each task to minimize cost and makepan.
\citea{442_shuPerformanceOptimizationHadoop2017} minimize makespan under a budget constraint by incorporating the fixed number of VM CPUs and their processing speed as variables.

VMs are typically allocated a limited, preconfigured amount of \concept{RAM}.
For example, \citea{442_shuPerformanceOptimizationHadoop2017} compare the RAM required for a task with the available working memory of a VM to ensure that the selected instance meets the task's memory demands.

Persistent \concept{Storage} is essential for data pipelines, for example, to transfer intermediate data between data centers as demonstrated by \citea{291_ikken_2018}.
While storage is often attached to a VM (e.g., \citea{83_nascimentoIncrementalReinforcementLearning2021}), some optimizations employ VM-independent storage services, such as AWS S3, as shown by \citea{334_quinnImplicationsAlternativeServerless2021} and \citea{424_fanOptimizingDataRegeneration2024}.

\paragraph{Other}

Two further parameters used in optimization cannot be assigned to the supercategories Cloud, Individual Tasks, and VM.
Thus, these parameters are presented in the fallback category \textit{Other}.

A further parameter is the \concept{past performance} of a system.
Optimizations can harness historical metrics -- such as makespan, the number of VMs deployed, and CPU utilization -- to forecast future performance.
For example, \citea{224_makhlouf2019data} derive these metrics from fully executed runs.
In contrast, \citea{182_aseman-manzarCostAwareResourceRecommendation2023} conduct reduced-scale executions of the data pipeline and to improve the cost-makespan ratio of the full pipeline.

The concept \concept{source code} encompasses strategies wherein the source code of data pipeline models plays an essential role in optimization beyond merely serving as a modeling tool.
For instance, \citea{194_dettiCWLPLASTaskWorkflows2023} extend the Common Workflow Language (CWL) with their custom extension CWL-PLAS to enable enhanced parallel computing.
Similarly, \citea{474_muller2014pydron} present the custom Python framework Pydron, which is designed to parallelize execution effectively.

\subsubsection{Algorithm Type} 
\label{subsubsec:realization_algorithmic_approach}

Defining an algorithm to optimize data pipelines is a complex task and is generally considered NP-hard \citep{291_ikken_2018, 604_chenTransformationBasedStreamingWorkflow2019, nezamiAnalysisEnergyEfficient2023a}. %
Therefore, the literature uses different classes of algorithms to approach their optimization method (see \autoref{tab:occurances_algorithmic_approach}).

\begin{table}[ht]
    \centering
    \footnotesize
    \caption{List of algorithm types used for optimizations}
    \label{tab:occurances_algorithmic_approach}
    \begin{tabularx}{\textwidth}{lrX}
        \toprule
        Algorithm & Count & Studies \\
        \midrule
Heuristic & 18 & \citepaperalias{19}, \citepaperalias{119}, \citepaperalias{593}, \citepaperalias{133}, \citepaperalias{604}, \citepaperalias{194}, \citepaperalias{450}, \citepaperalias{424}, \citepaperalias{291}, \citepaperalias{511}, \citepaperalias{379}, \citepaperalias{326}, \citepaperalias{474}, \citepaperalias{623}, \citepaperalias{442}, \citepaperalias{126}, \citepaperalias{426}, \citepaperalias{554} \\
ML/AI & 3 & \citepaperalias{449}, \citepaperalias{83}, \citepaperalias{229} \\
Simplified Model & 16 & \citepaperalias{182}, \citepaperalias{373}, \citepaperalias{593}, \citepaperalias{515}, \citepaperalias{291}, \citepaperalias{579}, \citepaperalias{322}, \citepaperalias{49}, \citepaperalias{511}, \citepaperalias{224}, \citepaperalias{35}, \citepaperalias{514}, \citepaperalias{334}, \citepaperalias{442}, \citepaperalias{487}, \citepaperalias{431} \\
        \bottomrule
    \end{tabularx}
\end{table}

\paragraph{Heuristics}
Heuristic algorithms approximate satisfactory solutions while keeping reasonable resource consumption \citea{Rardin2001ExperimentalEO}.
Among the various heuristic techniques, two notable examples are genetic algorithms and evolutionary algorithms.
\citea{424_fanOptimizingDataRegeneration2024} illustrate the potential of genetic algorithms for improved storage utilization and cost reduction, while \citea{326_mousavimojabICATSSchedulingBig2019} demonstrate the effectiveness of evolutionary algorithms to optimize scheduling decisions.
Importantly, these algorithms are not constrained to deterministic behavior, allowing for flexibility in reaching near-optimal solutions.

\paragraph{ML and AI}
Data pipelines can also be optimized using machine learning (ML) or artificial intelligence (AI).
Despite the general interest in ML/AI in recent years, only three studies from our study use this approach.
However, they are all relatively recent (from 2019, 2021, and 2023).
One RL approach is presented by \citea{83_nascimentoIncrementalReinforcementLearning2021}.
They employ WorkflowSim to simulate how pipelines would perform in a real cloud environment.
The simulation results enable their reinforcement learning algorithm to generate an optimized scheduling plan for actual execution in the cloud.

\paragraph{Simplified Model}
The real world is typically too complex to model directly, necessitating its abstraction into a simplified model.
By removing or combining less critical details, such an algorithm reduces the number of factors to consider, which in turn can lower the runtime.
While heuristic methods typically rely on statistical measures to formulate an execution plan, simplified models discard unimportant information to produce a plan within a reasonable time frame.
For example, linear programming techniques are employed in \citea{182_aseman-manzarCostAwareResourceRecommendation2023} and \citea{291_ikken_2018} to achieve this reduction in complexity.
In a different approach, \citea{511_kllapiScheduleOptimizationData2011} develop an approximation framework that generates execution schedules and benchmarks its performance against a generic greedy algorithm.

\section{Discussion} \label{sec:discussion}

In this section, we interpret the findings of the review.
We identify common patterns and pinpoint gaps in the current literature, which help guide future research directions in cloud-based data pipeline optimization and assist practitioners in refining their systems.

\subsection{Optimization Patterns}

Among the 33 studies analyzed, 27 incorporate scheduling optimizations to improve data pipeline performance.
However, scheduling is not always the primary focus, as only five studies rely solely on scheduling as their optimization method.
This suggests that while scheduling remains a crucial component of pipeline optimization, it is relatively well-explored in the literature.
Given this, we recommend that future research shifts focus toward alternative optimization methods, while still considering scheduling as an integral, but not exclusive, aspect of cloud-based data pipeline optimization.

The majority of studies analyzed focus on a single concept within each optimization context category, rather than exploring comparative approaches.
Specifically, 31 out of 33 studies consider only one data size, 29 out of 33 focus on a single processing paradigm, and 32 out of 33 examine only one type of infrastructure.
This reinforces the interpretation that most research explores isolated optimizations rather than comparative evaluations.
Furthermore, this pattern confirms that optimization context serves as a critical decision point for researchers designing data pipeline optimizations.

We observed that only three studies use ML/AI algorithms for their optimizations.
The limited adoption of these approaches can be attributed to several practical barriers, including higher computational overhead, the need for substantial training data, and the complexity of implementation and tuning.
These challenges are particularly difficult for pipeline systems that handle very diverse data pipelines with unique challenges.
In contrast, heuristic algorithms and simplified models are easier to implement and deploy yet still produce comparable results.
This explains their prevalence in the studies analyzed.

Several observed patterns serve as validations of the concept matrix.
For instance, we found that most studies discussing MapReduce-based optimization also consider data placement as a key optimization method.
This relationship is well-established in the literature \citep{wang2014dataLocalityMapReduce}, but its recurrence in our analysis further supports the matrix’s reliability.
Similarly, VM price is largely disregarded in studies optimizing for makespan, which aligns with the fact that makespan-focused optimizations do not inherently involve cost-related constraints.
These findings, while not novel, reinforce the internal consistency of the concept matrix within the scope of cloud-based data pipeline optimizations.

\subsection{Research Opportunities}

In terms of the context that current academic literature describes, we found a gap in studies addressing regular not big data.
Even for those studies that focus on \say{big data}, the definition of such remains unclear.
From the evaluations of those studies, we assume that most optimizations do not target data sizes in the terabyte area.
Nevertheless, we see value in exploring the regular data field, particularly from the angle of cloud vendors which might need to run large amounts of regular data pipelines.
The challenges then shift from optimizing a few large and long-running data pipelines to a plethora of smaller ones, leading to more dynamic load profiles and other optimization opportunities.
From the optimization methods we found, we see particular merit in exploring how caching between runs of a single pipeline in combination with data placement.

Most of the studies reviewed focus on optimizing single pipelines, highlighting a research gap in multi-tenant optimization techniques. This gap is particularly relevant given the increasing shift towards executing multiple parallel data pipelines, where challenges such as security, fair resource allocation, and workload balancing become more pronounced.
Despite the growing relevance of multi-pipeline execution models, most existing optimization techniques in our study selection are tailored to single-pipeline scenarios, overlooking the complex dependencies, scheduling conflicts, and dynamic resource provisioning required in multi-tenant systems.
Addressing this gap requires research of scalable, adaptive, and tenant-aware optimization approaches that balance efficiency, fairness, and security while maintaining cost-effectiveness in cloud-based pipeline infrastructures.

Another notable research gap is the lack of industry-based evaluations.
Workflows such as Montage \citep{jacob_2009_montage}, LIGO \citep{deeplman_2002_LIGO}, and Cybershake \citep{deelman_2006_cybershake} are commonly used to assess optimizations, providing a consistent benchmark for comparing performance across studies.
While these standardized workflows are valuable for ensuring reproducibility and controlled experimentation, real-world industry settings might differ in complexity, constraints, and unpredictable workloads.
As a result, many optimization techniques proposed in the literature may not be directly applicable or require significant adaptation before deployment in practical cloud environments.
Bridging this gap requires future research to incorporate industry-driven case studies, collaborations with cloud providers, and large-scale empirical evaluations to ensure that proposed optimizations are both theoretically sound and practically viable in real-world applications.

\section{Limitations \& Mitigations} \label{sec:limitations}

Our research design revolved around qualitative data analysis.
Thus, we used the trustworthiness criteria by \citet{guba1981criteria} to discuss the limitations of the study regarding credibility, transferability, dependability, and confirmability of the results.

\paragraph{Credibility} 
Our collection of primary materials presented a limit to how accurately the findings reflect reality.
We only reviewed literature that we found using the specified search terms, potentially excluding relevant literature that uses other terms for the same phenomenon.
We mitigated this effect by having conducted an ad-hoc pilot search to get familiar with the vocabulary of the domain. 
Because of that, among others, we decided to include the search term \textit{data workflows}.
Further, we excluded articles not written in English that might lead to missing relevant literature.
We believe this limitation is minor since English is the lingua franca of software engineering.
We only considered peer-reviewed articles to ensure a certain quality standard and trustworthiness of primary materials.
Gray literature might be more representative of industry-related data that we missed out on.
However, we took this trade-off deliberately to favor scientific rigor and data quality.

Besides our study selection, the overall body of knowledge might have been prone to publication bias.
Publications about failures are less likely to be published than success stories.
To accommodate a preferably broad spectrum of articles, we conducted our literature search on multiple established search engines, as well as on Google Scholar, which covers many more publication outlets.

During data analysis, we noticed that many studies did not fully describe the context of their optimizations.
For example, several studies did not report the infrastructures their optimizations are designed for, and the same for the processing paradigm.
Thus, we had to be more interpretative in those instances to classify the contexts of the studies.
To mitigate potential misinterpretations, we spent sufficient time with the data to infer the contexts reliably, but we explicitly used specific concepts to indicate this interpretative nature and included those in the reporting of the results.

\paragraph{Transferability}
While we expect our study to generalize well to individual data pipeline optimizations, we have to be careful assuming the same for multi-tenant service provides because further research would be needed.
The selected primary materials might not be representative of the broader population of interest.
For example, we did not find any study that explicitly proposes optimizations in a multi-tenant system, which we assume is typical for cloud providers.
Thus, further research is necessary to reliably make claims about how well the findings apply to multi-tenant environments.

Further, the lack of context descriptions in many studies and the use of the term \say{big data} in combination with the data sizes of evaluations raises doubts about the transferability of the findings to data pipelines of varying data sizes.
In this instance, further research is required to ensure generalizability for cloud systems working on a large amount of smaller data pipelines, as well as for data pipelines operating on large data sizes, e.g., with multiple terabytes of data.

\paragraph{Dependability}
Our literature selection and analysis are replicable only to a certain degree.
For example, the use of Google Scholar might be subject to geospatial differences in the search engine.
However, we transparently shared the underlying data of literature selection in our supplementary materials, so the selection process on the documented search results can be replicated.
The selection itself, using the inclusion and exclusion criteria, was not always clear in edge cases.
To mitigate this, we conducted the selection with multiple researchers in most parts, leading to investigator triangulation.
In parallel, the literature selection was prepared by the involved researchers, comparing results afterwards.
Conflicts were solved by discussion, feeding the insights back into refinements of the selection criteria to provide a replicable process.

The data analysis is a subjective process.
Although we paid attention to categorizing the results as objectively as possible, interpretations were unavoidable.
To mitigate this, we explicitly marked interpretations with a corresponding concept, and we conducted inter-coder agreement sessions.
Similar to the literature selection, multiple researchers categorized articles in parallel, and conflicts were resolved by discussion.
In this process, we reached agreement levels of $1.0$ for the literature selection and $0.85$ for the creation of the concept matrix, indicating strong to very strong inter-coder agreement among the researchers \citep{landisMeasurementObserverAgreement1977}.

\paragraph{Confirmability}
Our findings might be subject to biases on how the researchers' perspectives shaped the results.
This concerns the data analysis, mainly the more interpretative parts where we had to infer some elements of the context of an optimization.
To mitigate this risk of bias by the perspectives of a single researcher, we conducted the aforementioned inter-coder agreement sessions as investigator triangulation.

\section{Conclusion} \label{sec:conclusion}

This study explored optimization opportunities for cloud-based data pipeline infrastructures, focusing on application‐ and domain‐independent systems. Our work established a systematic framework that considers key contextual categories -- such as the optimization goal, data size, processing paradigm, and cloud infrastructure -- and links these to implementations like optimization methods, parameters, and algorithm types. In doing so, we found that cloud service providers can optimize the execution of multiple data pipelines by dynamically adopting techniques such as adaptive scheduling, provisioning, and efficient data transfer.

In addressing our first research question (RQ1), we demonstrated that a systematic approach incorporating contextual factors and tailored optimization implementations enables pipelines service providers to achieve improved execution efficiency. The analysis confirms that techniques such as dynamic scheduling, adaptive provisioning, and strategic data placement play critical roles in optimizing pipeline performance.
Furthermore, our concept matrix (\autoref{tab:matrix}) serves as a valuable tool for both practitioners and researchers by providing a clear framework to understand and apply the interrelationships among optimization goals, contexts, and implementations.

Our investigation of the specific optimization goals (RQ1.1) revealed seven primary objectives. These include cost minimization, makespan reduction, achieving minimal cost under makespan constraint, balancing cost-makespan trade-offs, minimizing makespan under a cost constraint, enhancing reactivity, and improving storage efficiency. Among these, makespan minimization and cost-makespan trade-offs emerged as the most prevalent goals.

For the last research question (RQ1.2), our systematic analysis identified scheduling as the most widely adopted optimization technique. This approach is complemented by provisioning, intra-task parallelism, data transfer optimization, and data placement. The majority of studies rely on heuristic algorithms and simplified models, with a lower but growing number exploring and ML/AI approaches.

Our review highlights notable research gaps, including the underexploration of multi-tenant environments and the optimization of smaller data pipelines, as well as the need for industry-driven case studies to validate these techniques in real-world settings. Overall, our findings offer a theoretical framework and practical guidance for future research and implementations in cloud-based data pipeline optimization.

\bibliographystyle{apalike}
\bibliography{arxiv}

\newpage
\appendix

\section{Study Mapping}
\label{tab:study_map}

\begin{table}[H]\centering
\caption{Mapping of Study IDs to Sources}
\begin{tabular}{llll}
\toprule
Study ID & Source & Author & Year \\
\midrule
A1\label{19_ahmadDynamicVMProvisioning2021} & \cite{19_ahmadDynamicVMProvisioning2021} & \citeauthor{19_ahmadDynamicVMProvisioning2021} & \citeyear{19_ahmadDynamicVMProvisioning2021} \\
A2\label{182_aseman-manzarCostAwareResourceRecommendation2023} & \cite{182_aseman-manzarCostAwareResourceRecommendation2023} & \citeauthor{182_aseman-manzarCostAwareResourceRecommendation2023} & \citeyear{182_aseman-manzarCostAwareResourceRecommendation2023} \\
A3\label{373_changxin_lpod} & \cite{373_changxin_lpod} & \citeauthor{373_changxin_lpod} & \citeyear{373_changxin_lpod} \\
A4\label{119_bousselmi_bi_objective} & \cite{119_bousselmi_bi_objective} & \citeauthor{119_bousselmi_bi_objective} & \citeyear{119_bousselmi_bi_objective} \\
A5\label{593_bugingoDecompositionBasedMultiobjective2021} & \cite{593_bugingoDecompositionBasedMultiobjective2021} & \citeauthor{593_bugingoDecompositionBasedMultiobjective2021} & \citeyear{593_bugingoDecompositionBasedMultiobjective2021} \\
A6\label{133_chenBigDataProcessing2020} & \cite{133_chenBigDataProcessing2020} & \citeauthor{133_chenBigDataProcessing2020} & \citeyear{133_chenBigDataProcessing2020} \\
A7\label{449_chenPISCESOptimizingMultiJob2019} & \cite{449_chenPISCESOptimizingMultiJob2019} & \citeauthor{449_chenPISCESOptimizingMultiJob2019} & \citeyear{449_chenPISCESOptimizingMultiJob2019} \\
A8\label{604_chenTransformationBasedStreamingWorkflow2019} & \cite{604_chenTransformationBasedStreamingWorkflow2019} & \citeauthor{604_chenTransformationBasedStreamingWorkflow2019} & \citeyear{604_chenTransformationBasedStreamingWorkflow2019} \\
A9\label{194_dettiCWLPLASTaskWorkflows2023} & \cite{194_dettiCWLPLASTaskWorkflows2023} & \citeauthor{194_dettiCWLPLASTaskWorkflows2023} & \citeyear{194_dettiCWLPLASTaskWorkflows2023} \\
A10\label{515_ebrahimiSchedulingBigData2018} & \cite{515_ebrahimiSchedulingBigData2018} & \citeauthor{515_ebrahimiSchedulingBigData2018} & \citeyear{515_ebrahimiSchedulingBigData2018} \\
A11\label{450_esteves_2014a} & \cite{450_esteves_2014a} & \citeauthor{450_esteves_2014a} & \citeyear{450_esteves_2014a} \\
A12\label{424_fanOptimizingDataRegeneration2024} & \cite{424_fanOptimizingDataRegeneration2024} & \citeauthor{424_fanOptimizingDataRegeneration2024} & \citeyear{424_fanOptimizingDataRegeneration2024} \\
A13\label{291_ikken_2018} & \cite{291_ikken_2018} & \citeauthor{291_ikken_2018} & \citeyear{291_ikken_2018} \\
A14\label{579_kacsukFlowbsterCloudOrientedWorkflow2018} & \cite{579_kacsukFlowbsterCloudOrientedWorkflow2018} & \citeauthor{579_kacsukFlowbsterCloudOrientedWorkflow2018} & \citeyear{579_kacsukFlowbsterCloudOrientedWorkflow2018} \\
A15\label{322_kamburugamuveHPTMTOperatorBasedArchitecture2021} & \cite{322_kamburugamuveHPTMTOperatorBasedArchitecture2021} & \citeauthor{322_kamburugamuveHPTMTOperatorBasedArchitecture2021} & \citeyear{322_kamburugamuveHPTMTOperatorBasedArchitecture2021} \\
A16\label{49_kashlevSystemArchitectureRunning2014} & \cite{49_kashlevSystemArchitectureRunning2014} & \citeauthor{49_kashlevSystemArchitectureRunning2014} & \citeyear{49_kashlevSystemArchitectureRunning2014} \\
A17\label{511_kllapiScheduleOptimizationData2011} & \cite{511_kllapiScheduleOptimizationData2011} & \citeauthor{511_kllapiScheduleOptimizationData2011} & \citeyear{511_kllapiScheduleOptimizationData2011} \\
A18\label{224_makhlouf2019data} & \cite{224_makhlouf2019data} & \citeauthor{224_makhlouf2019data} & \citeyear{224_makhlouf2019data} \\
A19\label{379_mehranMatchingbasedSchedulingAsynchronous2022} & \cite{379_mehranMatchingbasedSchedulingAsynchronous2022} & \citeauthor{379_mehranMatchingbasedSchedulingAsynchronous2022} & \citeyear{379_mehranMatchingbasedSchedulingAsynchronous2022} \\
A20\label{35_mohanNoSQLDataModel2016} & \cite{35_mohanNoSQLDataModel2016} & \citeauthor{35_mohanNoSQLDataModel2016} & \citeyear{35_mohanNoSQLDataModel2016} \\
A21\label{514_aravind_schedulingBigData} & \cite{514_aravind_schedulingBigData} & \citeauthor{514_aravind_schedulingBigData} & \citeyear{514_aravind_schedulingBigData} \\
A22\label{326_mousavimojabICATSSchedulingBig2019} & \cite{326_mousavimojabICATSSchedulingBig2019} & \citeauthor{326_mousavimojabICATSSchedulingBig2019} & \citeyear{326_mousavimojabICATSSchedulingBig2019} \\
A23\label{474_muller2014pydron} & \cite{474_muller2014pydron} & \citeauthor{474_muller2014pydron} & \citeyear{474_muller2014pydron} \\
A24\label{83_nascimentoIncrementalReinforcementLearning2021} & \cite{83_nascimentoIncrementalReinforcementLearning2021} & \citeauthor{83_nascimentoIncrementalReinforcementLearning2021} & \citeyear{83_nascimentoIncrementalReinforcementLearning2021} \\
A25\label{334_quinnImplicationsAlternativeServerless2021} & \cite{334_quinnImplicationsAlternativeServerless2021} & \citeauthor{334_quinnImplicationsAlternativeServerless2021} & \citeyear{334_quinnImplicationsAlternativeServerless2021} \\
A26\label{623_hamid_workflowScheduling} & \cite{623_hamid_workflowScheduling} & \citeauthor{623_hamid_workflowScheduling} & \citeyear{623_hamid_workflowScheduling} \\
A27\label{442_shuPerformanceOptimizationHadoop2017} & \cite{442_shuPerformanceOptimizationHadoop2017} & \citeauthor{442_shuPerformanceOptimizationHadoop2017} & \citeyear{442_shuPerformanceOptimizationHadoop2017} \\
A28\label{126_simicBigDataBPMN2023} & \cite{126_simicBigDataBPMN2023} & \citeauthor{126_simicBigDataBPMN2023} & \citeyear{126_simicBigDataBPMN2023} \\
A29\label{487_tranReducingTailLatencies2018} & \cite{487_tranReducingTailLatencies2018} & \citeauthor{487_tranReducingTailLatencies2018} & \citeyear{487_tranReducingTailLatencies2018} \\
A30\label{431_tudoranOverFlowMultiSiteAware2016} & \cite{431_tudoranOverFlowMultiSiteAware2016} & \citeauthor{431_tudoranOverFlowMultiSiteAware2016} & \citeyear{431_tudoranOverFlowMultiSiteAware2016} \\
A31\label{426_wuOptimizingPerformanceBig2016} & \cite{426_wuOptimizingPerformanceBig2016} & \citeauthor{426_wuOptimizingPerformanceBig2016} & \citeyear{426_wuOptimizingPerformanceBig2016} \\
A32\label{554_yeStorageAwareTaskScheduling2018} & \cite{554_yeStorageAwareTaskScheduling2018} & \citeauthor{554_yeStorageAwareTaskScheduling2018} & \citeyear{554_yeStorageAwareTaskScheduling2018} \\
A33\label{229_zhangDataintensiveWorkflowScheduling2023} & \cite{229_zhangDataintensiveWorkflowScheduling2023} & \citeauthor{229_zhangDataintensiveWorkflowScheduling2023} & \citeyear{229_zhangDataintensiveWorkflowScheduling2023} \\
\bottomrule
\end{tabular}

\end{table} 
\newpage

\section{Concept Matrix}
\label{tab:matrix}

The following tables contain the code matrix we created for data analysis.
For presentation, we divided the code matrix into multiple tables for better readability.

\newcommand{\conceptmatrixheadercol}[1]{%
    \multicolumn{1}{l}{\parbox[t]{0.3cm}{\centering \rotatebox{45}{#1}}}%
}

\begin{table}[H]
\centering
\caption{Concept Matrix---Processing Paradigm, Data Size, Infrastructure, and Algorithm Type}
\footnotesize
\begin{tabular}{l|cccc|cc|cccc|ccc}
    
    \multicolumn{1}{l}{} 
    & \conceptmatrixheadercol{Not explicitly mentioned}
    & \conceptmatrixheadercol{Batch}
    & \conceptmatrixheadercol{MapReduce}
    & \conceptmatrixheadercol{Stream}
    & \conceptmatrixheadercol{Big Data}
    & \conceptmatrixheadercol{Regular Data}
    & \conceptmatrixheadercol{Not explicitly mentioned}
    & \conceptmatrixheadercol{Multi-Cloud}
    & \conceptmatrixheadercol{Single-Cloud (Heterogeneous)}
    & \conceptmatrixheadercol{Single-Cloud (Homogeneous)} 
    & \conceptmatrixheadercol{Heuristic}
    & \conceptmatrixheadercol{ML/AI}
    & \conceptmatrixheadercol{Simplified Model}
    \\

    \multicolumn{1}{l}{Study}& \multicolumn{4}{c}{Processing Paradigm} & \multicolumn{2}{c}{Data Size} & \multicolumn{4}{c}{Infrastructure} & \multicolumn{3}{c}{Algorithm Type} \\

    \midrule
    \citepaperalias{19} 
        & X & X & . & . 
        & X & . 
        & . & . & X & . 
        & X & . & . \\
    \citepaperalias{182} 
        & X & X & . & . 
        & X & . 
        & . & . & X & . 
        & . & . & X \\
    \citepaperalias{373} 
        & X & X & . & . 
        & X & . 
        & . & . & X & . 
        & . & . & X \\
    \citepaperalias{119} 
        & X & X & . & . 
        & X & . 
        & X & . & X & . 
        & X & . & . \\
    \citepaperalias{593} 
        & X & X & . & . 
        & X & . 
        & X & . & X & . 
        & X & . & X \\
    \citepaperalias{133} 
        & X & . & . & X 
        & X & . 
        & . & X & . & . 
        & X & . & . \\
    \citepaperalias{449} 
        & . & . & X & . 
        & X & . 
        & X & . & . & X 
        & . & X & . \\
    \citepaperalias{604} 
        & . & . & . & X 
        & X & . 
        & . & X & . & . 
        & X & . & . \\
    \citepaperalias{194} 
        & X & X & . & . 
        & X & . 
        & X & . & X & . 
        & X & . & . \\
    \citepaperalias{515} 
        & X & X & . & . 
        & X & . 
        & X & . & X & . 
        & . & . & X \\
    \citepaperalias{450} 
        & . & . & . & X 
        & X & . 
        & X & . & X & . 
        & X & . & . \\
    \citepaperalias{424} 
        & X & X & . & . 
        & X & X 
        & . & . & . & X 
        & X & . & . \\
    \citepaperalias{291} 
        & X & X & . & . 
        & X & . 
        & . & X & . & . 
        & X & . & X \\
    \citepaperalias{579} 
        & . & . & . & X 
        & X & . 
        & X & X & X & . 
        & . & . & X \\
    \citepaperalias{322} 
        & . & X & . & X 
        & X & . 
        & X & . & . & X 
        & . & . & X \\
    \citepaperalias{49} 
        & X & X & . & . 
        & X & . 
        & X & . & X & . 
        & . & . & X \\
    \citepaperalias{511} 
        & . & X & . & X 
        & X & . 
        & . & . & . & X 
        & X & . & X \\
    \citepaperalias{224} 
        & X & X & . & . 
        & X & . 
        & X & . & . & X 
        & . & . & X \\
    \citepaperalias{379} 
        & X & . & . & X 
        & . & X 
        & . & X & . & . 
        & X & . & . \\
    \citepaperalias{35} 
        & . & . & X & . 
        & X & . 
        & X & . & . & X 
        & . & . & X \\
    \citepaperalias{514} 
        & X & X & . & . 
        & X & . 
        & X & . & X & . 
        & . & . & X \\
    \citepaperalias{326} 
        & X & X & . & . 
        & X & . 
        & X & . & X & . 
        & X & . & . \\
    \citepaperalias{474} 
        & X & X & . & . 
        & . & X 
        & X & . & X & . 
        & X & . & . \\
    \citepaperalias{83} 
        & X & X & . & . 
        & X & . 
        & X & . & X & . 
        & . & X & . \\
    \citepaperalias{334}
        & . & X & . & . 
        & . & X 
        & . & . & X & . 
        & . & . & X \\
    \citepaperalias{623} 
        & X & X & . & . 
        & X & . 
        & X & . & X & . 
        & X & . & . \\
    \citepaperalias{442} 
        & X & X & X & . 
        & X & . 
        & . & . & X & . 
        & X & . & X \\
    \citepaperalias{126} 
        & . & X & . & . 
        & X & . 
        & . & . & X & . 
        & X & . & . \\
    \citepaperalias{487} 
        & . & . & . & X 
        & X & . 
        & X & . & . & X 
        & . & . & X \\
    \citepaperalias{431} 
        & . & X & X & . 
        & X & . 
        & . & X & . & . 
        & . & . & X \\
    \citepaperalias{426} 
        & X & X & . & . 
        & X & X 
        & . & X & . & . 
        & X & . & . \\
    \citepaperalias{554} 
        & . & . & X & . 
        & X & . 
        & X & . & X & . 
        & X & . & . \\
    \citepaperalias{229} 
        & X & X & . & . 
        & X & . 
        & . & X & . & . 
        & . & X & . \\
\end{tabular}
\end{table}

\begin{table}[H]
    \centering
    \caption{Concept Matrix---Goal and Method}
    \footnotesize
    \begin{tabular}{l|ccccccc|cccccccccc}
    \multicolumn{1}{c}{} 
    & \conceptmatrixheadercol{Cost}
    & \conceptmatrixheadercol{Cost Under Makespan Constraint}
    & \conceptmatrixheadercol{Cost-Makespan-Tradeoff}
    & \conceptmatrixheadercol{Makespan}
    & \conceptmatrixheadercol{Makespan Under Cost Constraint}
    & \conceptmatrixheadercol{Reactivity}
    & \conceptmatrixheadercol{Storage}
    & \conceptmatrixheadercol{Caching}
    & \conceptmatrixheadercol{Data Placement}
    & \conceptmatrixheadercol{Data Regeneration}
    & \conceptmatrixheadercol{Data Transfer}
    & \conceptmatrixheadercol{Execution Overlapping}
    & \conceptmatrixheadercol{Intra-Task Parallelism}
    & \conceptmatrixheadercol{Model Transformation}
    & \conceptmatrixheadercol{Provisioning}
    & \conceptmatrixheadercol{Redundancy}
    & \conceptmatrixheadercol{Scheduling}
    \\
    \multicolumn{1}{l}{Study} & \multicolumn{7}{c}{Goal} & \multicolumn{10}{c}{Method} \\

    \midrule

    \citepaperalias{19}
        & . & X & . & . & . & . & . 
        & . & . & . & . & . & . & X & X & . & X 
        \\\
    \citepaperalias{182}
        & . & . & X & . & . & . & . 
        & . & . & . & . & . & . & . & X & . & X 
        \\
    \citepaperalias{373}
        & . & X & . & . & . & . & . 
        & . & . & . & . & . & . & . & X & . & X 
        \\
    \citepaperalias{119}
        & . & . & X & . & . & . & . 
        & . & . & . & . & . & . & . & . & . & X 
        \\
    \citepaperalias{593}
        & . & . & X & . & . & . & . 
        & . & . & . & . & . & . & . & . & . & X 
        \\
    \citepaperalias{133}
        & . & . & X & . & . & . & . 
        & . & . & . & . & . & . & . & X & X & X 
        \\
    \citepaperalias{449}
        & . & . & . & X & . & . & . 
        & . & X & . & . & X & . & . & . & . & X 
        \\
    \citepaperalias{604}
        & X & . & . & . & . & . & . 
        & . & X & . & X & . & . & X & . & . & X 
        \\
    \citepaperalias{194}
        & . & . & . & X & . & . & . 
        & . & . & . & . & . & X & . & . & . & X 
        \\
    \citepaperalias{515}
        & . & X & . & . & . & . & . 
        & . & . & . & . & . & . & . & X & . & X 
        \\
    \citepaperalias{450}
        & . & X & . & . & . & . & . 
        & . & . & . & . & . & . & X & X & . & X 
        \\
    \citepaperalias{424}
        & X & . & . & . & . & . & X 
        & . & . & X & . & . & . & . & . & . & . 
        \\
    \citepaperalias{291}
        & X & . & . & . & . & . & X 
        & . & X & . & . & . & X & . & . & . & . 
        \\
    \citepaperalias{579}
        & . & . & . & X & . & . & . 
        & . & . & . & . & X & X & X & X & . & X 
        \\
    \citepaperalias{322}
        & . & . & . & X & . & . & . 
        & . & . & . & . & . & X & . & . & . & . 
        \\
    \citepaperalias{49}
        & . & . & X & X & . & . & . 
        & . & . & . & X & . & . & . & . & . & X 
        \\
    \citepaperalias{511}
        & . & X & X & . & X & . & . 
        & . & . & . & X & . & . & . & . & . & X 
        \\
    \citepaperalias{224}
        & X & . & . & . & . & . & . 
        & . & X & . & X & . & . & . & X & . & X 
        \\
    \citepaperalias{379}
        & . & . & . & X & . & . & . 
        & . & . & . & . & . & . & . & . & . & X 
        \\
    \citepaperalias{35}
        & . & . & . & X & . & . & . 
        & . & X & . & . & . & X & . & X & . & X 
        \\
    \citepaperalias{514}
        & . & . & . & . & X & . & . 
        & . & . & . & . & . & . & . & X & . & X 
        \\
    \citepaperalias{326}
        & . & X & . & . & . & . & . 
        & . & . & . & . & . & . & . & X & . & X 
        \\
    \citepaperalias{474}
        & . & . & . & X & . & . & . 
        & . & . & . & . & . & X & X & . & . & X 
        \\
    \citepaperalias{83}
        & . & . & . & X & . & . & . 
        & . & . & . & . & . & X & . & . & . & X 
        \\
    \citepaperalias{334}
        & . & . & X & . & . & . & . 
        & . & . & . & . & . & . & . & . & . & X 
        \\
    \citepaperalias{623}
        & . & . & . & X & . & . & . 
        & . & . & . & . & . & . & . & . & . & X 
        \\
    \citepaperalias{442}
        & . & . & . & . & X & . & . 
        & . & . & . & . & . & X & . & . & . & X 
        \\
    \citepaperalias{126}
        & . & . & X & . & . & . & . 
        & . & . & . & . & . & . & . & X & . & . 
        \\
    \citepaperalias{487}
        & . & . & . & X & . & X & . 
        & . & . & . & . & . & . & . & . & X & . 
        \\
    \citepaperalias{431}
        & . & . & X & . & . & . & . 
        & X & X & . & X & . & . & . & . & . & . 
        \\
    \citepaperalias{426}
        & . & . & . & . & X & . & . 
        & . & . & . & X & . & . & . & X & . & X 
        \\
    \citepaperalias{554}
        & . & . & . & X & . & . & . 
        & . & X & . & X & . & X & . & . & . & X 
        \\
    \citepaperalias{229}
        & . & . & . & X & . & . & . 
        & . & . & . & X & . & . & X & . & . & X 
        \\
    \end{tabular}
\end{table}

\begin{table}[H]
\centering
\caption{Concept Matrix---Parameters}
\footnotesize
\begin{tabular}{l|ccccccccccccccc}
\multicolumn{1}{c}{} 
& \conceptmatrixheadercol{Cloud / Data Transfer Price} 
& \conceptmatrixheadercol{Cloud / Data Transfer Rate} 
& \conceptmatrixheadercol{Cloud / VM-Types} 
& \conceptmatrixheadercol{Task / Data Size} 
& \conceptmatrixheadercol{Task / Dependencies} 
& \conceptmatrixheadercol{Task / RAM} 
& \conceptmatrixheadercol{Task / CPU} 
& \conceptmatrixheadercol{Past Performance} 
& \conceptmatrixheadercol{Source Code} 
& \conceptmatrixheadercol{VM / Billing Cycle} 
& \conceptmatrixheadercol{VM / Boot Delay} 
& \conceptmatrixheadercol{VM / CPU} 
& \conceptmatrixheadercol{VM / Price} 
& \conceptmatrixheadercol{VM / RAM} 
& \conceptmatrixheadercol{VM / Storage} 
\\
\multicolumn{1}{l}{Study} & \multicolumn{15}{c}{Parameters} \\

\midrule
\citepaperalias{19} & . & X & X & . & X & . & . & . & . & X & X & . & X & . & . \\
\citepaperalias{182} & . & . & X & . & X & . & . & X & . & . & . & . & X & . & . \\
\citepaperalias{373} & X & X & X & X & X & . & X & X & . & X & X & X & X & . & . \\
\citepaperalias{119} & X & X & . & X & X & . & X & . & . & . & . & . & X & . & . \\
\citepaperalias{593} & . & X & X & . & X & . & X & . & . & X & . & X & X & . & . \\
\citepaperalias{133} & . & X & X & X & X & . & . & . & . & X & . & . & X & . & . \\
\citepaperalias{449} & . & . & . & X & X & . & X & . & . & . & . & . & . & . & . \\
\citepaperalias{604} & X & . & . & . & X & . & . & . & . & . & . & X & X & . & . \\
\citepaperalias{194} & . & . & . & . & X & . & . & . & X & . & . & . & . & . & . \\
\citepaperalias{515} & X & X & X & X & X & . & X & . & . & . & . & X & X & . & . \\
\citepaperalias{450} & . & . & X & . & . & X & X & X & . & X & . & . & X & . & . \\
\citepaperalias{424} & . & . & . & . & X & . & . & . & . & . & . & X & X & . & X \\
\citepaperalias{291} & X & X & . & X & X & . & . & . & . & . & . & . & X & . & X \\
\citepaperalias{579} & . & . & X & . & X & . & . & . & X & . & . & . & . & . & . \\
\citepaperalias{322} & . & . & . & . & . & . & . & . & X & . & . & . & . & . & . \\
\citepaperalias{49} & . & . & X & . & . & . & . & . & . & . & . & . & . & . & . \\
\citepaperalias{511} & . & X & . & . & X & X & X & . & . & X & . & X & X & X & . \\
\citepaperalias{224} & . & X & . & X & X & . & X & X & . & X & . & . & X & . & . \\
\citepaperalias{379} & . & X & . & X & X & X & X & . & . & . & . & X & . & X & X \\
\citepaperalias{35} & . & . & . & X & X & . & . & . & . & . & . & . & . & . & . \\
\citepaperalias{514} & . & X & X & X & X & . & X & X & . & . & . & . & X & . & . \\
\citepaperalias{326} & X & X & X & X & X & . & X & . & . & . & . & X & X & . & . \\
\citepaperalias{474} & . & . & . & . & . & . & . & . & X & . & . & . & . & . & . \\
\citepaperalias{83} & . & X & X & . & X & . & X & X & . & . & . & X & X & . & X \\
\citepaperalias{334} & . & . & . & . & . & . & . & . & . & . & X & X & X & X & X \\
\citepaperalias{623} & . & X & X & . & X & . & . & . & . & . & . & . & . & . & . \\
\citepaperalias{442} & . & . & X & . & X & X & X & . & . & X & . & X & X & X & . \\
\citepaperalias{126} & . & . & X & . & . & . & X & . & . & . & X & . & X & . & . \\
\citepaperalias{487} & . & . & . & . & X & . & . & . & . & . & . & . & . & . & . \\
\citepaperalias{431} & X & X & . & X & X & . & . & X & . & . & . & X & X & X & X \\
\citepaperalias{426} & X & X & X & X & X & . & . & . & . & X & . & X & X & X & X \\
\citepaperalias{554} & . & X & X & X & X & X & X & . & . & . & . & X & . & X & X \\
\citepaperalias{229} & . & X & . & . & X & . & . & . & . & . & . & X & . & . & . \\
\end{tabular}
\end{table}

\end{document}